\documentclass[12pt]{iopart}
\usepackage{iopams,graphicx,color}
\begin{document}
\def\be{\begin{equation}}
\def\ee{\end{equation}}
\def\bea{\begin{eqnarray}}
\def\eea{\end{eqnarray}}
\def\theta{\vartheta}
\title{Surprises from quenches in long-range interacting systems: Temperature inversion and cooling}
\author{Shamik Gupta$^1$\footnote{Corresponding author; Email: shamikg1@gmail.com}, Lapo Casetti$^{2,3}$\footnote{Email: lapo.casetti@unifi.it}}
$^1$ Max-Planck-Institut f\"{u}r Physik Komplexer Systeme, N\"{o}thnitzer Stra\ss e 38, D-01187 Dresden, Germany \\
$^2$ Dipartimento di Fisica e Astronomia and CSDC, Universit\`a di
Firenze, and INFN, sezione di Firenze, via G.\ Sansone 1, I-50019 Sesto
Fiorentino, Italy \\
$^3$ INAF-Osservatorio Astrofisico di Arcetri, Largo E.\ Fermi 5, I-50125 Firenze, Italy
\begin{abstract}
What happens when one of the parameters governing the dynamics of a long-range interacting system of particles in thermal equilibrium is abruptly changed (quenched) to a different value?
While a short-range system, under the same conditions, will relax in
time to a new thermal equilibrium with a uniform temperature across the
system, a long-range system shows a fast relaxation to a nonequilibrium
{\em quasistationary state} (QSS). The lifetime of such an
off-equilibrium state diverges with the system size, and the temperature
is non-uniform across the system. Quite surprisingly, the density
profile in the QSS obtained after the quench is anticorrelated with the
temperature profile in space, thus exhibiting the phenomenon of {\em
temperature inversion}: denser regions are colder than sparser ones.  We
illustrate with extensive molecular dynamics simulations the ubiquity of
this scenario in a prototypical long-range interacting system subject to
a variety of quenching protocols, and in a model that mimics an
experimental setup of atoms interacting with light in an optical cavity.
We further demonstrate how a procedure of iterative quenching combined
with filtering out the high-energy particles in the system may be
employed to cool the system. Temperature inversion is observed in nature
in some astrophysical settings; our results imply that such a phenomenon
should be observable, and could even be exploitable to advantage, also in controlled laboratory experiments. 
\end{abstract}
\date{\today}
\maketitle
\section{Introduction}
Long-range-interacting systems abound in nature
\cite{Campa:2009,Bouchet:2010,Campa:2014}.
Interactions are called long-range when the pair potential energy decays asymptotically with the interparticle distance $r$ as $r^{-\alpha}$, with $0 \le \alpha \le
d$ in $d$ spatial dimensions, as it happens for instance in Coulomb and
gravitational interactions. One striking feature of many-particle
long-range interacting systems is that they are typically found in
states that are out of thermodynamic equilibrium. This is at variance
with systems with short-range interactions: while nonequilibrium states
in short-range systems require an enduring external forcing to
counteract the tendency of the system to reach due to ``collisional''
effects a thermal equilibrium state, an isolated long-range interacting
system evolving while starting far from equilibrium will remain stuck in
an out-of-equilibrium state for very long times. Such times grow with
the number $N$ of degrees of freedom, so that they may become larger
than any experimentally accessible timescale. For instance, the time
needed for a galaxy to reach an ``equilibrium'' state\footnote{For
non-confined self-gravitating systems with a finite mass, the ``true''
thermodynamic equilibrium state where the velocities obey a
Maxwell-Boltzmann distribution is never reached; this estimate refers to the time needed to reach a state where the memory of the initial condition has been completely lost \protect\cite{Choudouri:2010,Binney:2008}.} is estimated to be several orders of magnitude larger than the age of the universe \cite{Choudouri:2010,Binney:2008}. 
This property is universal, in the sense that it is shared by all
systems with long-range interactions, regardless of the details of the
interaction potential \footnote{This is true provided there are only
long-range interactions; this property does not hold for systems with mixed long- and short-range interactions.}. It is understood as a consequence of the fact that the kinetic equation governing the evolution of the single-particle distribution function $f(\mathbf{q},\mathbf{p},t)$, where $\mathbf{q}$ and $\mathbf{p}$ are the canonically conjugated positions and momenta, respectively, is the Vlasov equation (often referred to as the collisionless Boltzmann equation, mainly in the astrophysical literature), up to a timescale $\tau_{\mathrm{coll}}$ when collisional effects can no longer be neglected, with $\tau_{\mathrm{coll}}$ diverging with $N$ \cite{Campa:2009,Campa:2014}. Such an equation has infinitely many stationary solutions: among these, the stable ones are identified with the observed non-equilibrium states, often referred to as {\em quasi-stationary states} (QSSs), and the thermal equilibrium state is just one out of infinitely many. 

The overall qualitative picture of the dynamical evolution of an
isolated long-range system starting far from thermal equilibrium is the
following: After a transient (referred to as {\em violent relaxation}
after Lynden-Bell \cite{Lynden-Bell:1967}) whose lifetime does not
depend on $N$, the system sets into a QSS and stays there for a time of
the order of $\tau_{\mathrm{coll}}$. For times $t >
\tau_{\mathrm{coll}}$, when the Vlasov description no longer applies,
the QSS is no longer stationary, and the system eventually evolves to a
thermal equilibrium state. The QSS the system relaxes to after the
violent relaxation depends on the initial conditions, and no general
theory is available yet to predict it. A completely general statistical
approach was proposed by Lynden-Bell \cite{Lynden-Bell:1967}, but it
rests on the hypothesis of complete mixing of the Vlasov dynamics that
only rarely holds, hence, gives reasonably accurate predictions only in
very particular cases \cite{Campa:2014,Levin:2014}. Other methods have
been proposed since then, which give good predictions for simple models only for special classes of initial conditions, namely, for single-level initial distributions (the so-called ``water-bag'' distributions) \cite{Campa:2014,Levin:2014} or when the relaxation towards the QSS is nearly adiabatic \cite{Levin:2014,Benetti:2014,Leoncini:2009,DeBuyl:2011}. 

Although predicting the full distribution function in a QSS resulting
from a generic initial condition is a formidable task, especially for
spatially inhomogeneous QSSs whose stability properties are particularly
difficult to analyze, one may ask whether the QSSs that are reached in
certain situations do share some common features. One of the hallmarks
of thermal equilibrium is a uniform temperature throughout the system,
hence, also in inhomogeneous equilibrium states, the temperature will be
constant. Conversely, in a nonequilibrium state, the temperature may well be
nonuniform. Hence, one may ask whether some properties of the temperature profile in a QSS share a certain degree of universality. 
A physically relevant case is that of a system prepared in a (spatially
inhomogeneous) thermal equilibrium state and then brought out of
equilibrium by means of a (non-small) sudden perturbation, acting for a
very short time. As shown in recent papers \cite{Teles:2015,Teles:2016},
in this case, the typical outcome is somewhat surprising and
counterintuitive, and is referred to as {\em temperature inversion}. The
latter effect implies that the density and temperature profiles are
anticorrelated: sparser regions of the system are hotter than denser
ones. Temperature inversions are observed in nature in astrophysical settings, the most famous example being the solar corona,
where the temperature grows from thousands to millions of Kelvin while
going from the photosphere to the sparser external regions of the corona
\cite{GolubPasachoff:2009}. Other examples of temperature inversions
have been observed, for instance, in molecular clouds \cite{Myers:1992}. As argued in \cite{Teles:2015,Teles:2016,Casetti:2014},
temperature inversion should not be a peculiarity of systems in
extreme conditions, as astrophysical systems typically are, but should
rather be a generic property of long-range interacting systems relaxing
to a QSS after having been brought out of (spatially inhomogeneous)
thermal equilibrium by a perturbation. In \cite{Teles:2015}, temperature
inversion was observed as a result of perturbations of a thermal state of a
prototypical model with long-range interactions, the Hamiltonian
mean-field (HMF) model \cite{Antoni:1995}, and of a two-dimensional
self-gravitating system. Temperature inversion in the latter system is
thoroughly investigated in \cite{Cintio:2016} in connection with
filamentary structures in galactic molecular clouds. Partial temperature
inversions were also observed in QSSs of two-dimensional
self-gravitating systems whose initial conditions were of the water-bag
type \cite{Teles:2010}. In one-dimensional self-gravitating systems,
temperature inversions in QSSs were observed, which gradually
disappear during the slow relaxation of the system from the QSS to
thermal equilibrium \cite{W_thesis}. Temperature inversion was recently demonstrated to occur in the nonequilibrium stationary state of a class of mean-field systems
involving rotators subject to quenched disordered external drive and
dissipation \cite{Campa:2015}. A physical picture of the origin of temperature inversion in a generic long-range interacting system, based on the interplay between a mechanism originally proposed to explain the temperature profile in the solar corona  (referred to as {\em velocity filtration} \cite{Scudder:1992-1,Scudder:1992-2,Scudder:1994,supplmat:2015}) and the interaction of the system particles with the time-dependent mean field, was also suggested \cite{Teles:2015}.

If the phenomenon of temperature inversion is somewhat universal in
long-range interacting systems abruptly brought out of thermal
equilibrium, one may wonder whether it could be observed also in a
controlled laboratory experiment. In the present paper, we aim at taking a step forward towards a positive answer to this question. 
First of all, using molecular dynamics (MD) simulations, we demonstrate
that in the HMF model, with both attractive and repulsive interactions,
temperature inversion is typically observed when the system is brought
out of thermal equilibrium by means of an abrupt change (from now on
referred to as a {\em quench}) of a parameter controlling the dynamics
of the system, like a coupling constant or an external field. To this
end, we employ a variety of different quench protocols. Our studies are
relevant to experiments because quench protocols can be performed with
experimental setups, especially in systems of (cold) atoms; moreover, an
HMF model with repulsive interactions and an external confining field
can be seen as a toy model of trapped ions, while an HMF model with
attractive interactions does share many features with atoms interacting
with light in a cavity. We then elaborate on this point by demonstrating
that at least one of the considered quench protocols can be applied to
another model, which mimics more closely an experimental setup of atoms interacting with a standing electromagnetic wave in an optical cavity \cite{Schutz:2014,Schutz:2015,Jager:2016}, again yielding temperature inversion. It is worth noting that the quenching protocol applicable to the latter model does correspond to changing a parameter that can be actually tuned in an experiment, that is, the intensity of an external laser pump. We also give evidence that the physical picture proposed in \cite{Teles:2015} does apply to temperature inversions produced by quenches. Finally, we show that an iterative quenching protocol, combined with filtering out the high-energy particles localized in the sparser external regions, can be used to cool the system.

The paper is organized as follows: in Sec.\ \ref{sec:HMF}, we introduce
the HMF model, and study the temperature inversion produced by quenching
the external field in the case of repulsive interactions
(\S\ref{sec:AF-qh}) as well as that produced in the case with attractive
interactions by quenching the external field (\S\ref{sec:F-qh}) and the
coupling constant (\S\ref{sec:F-qJ}). In Sec.\ \ref{sec:Morigi}, we
discuss a model of atoms interacting with light in an optical cavity,
with emphasis on the dissipationless limit (\S\ref{sec:dissipless}). We
then report on temperature inversion as observed in the latter system
after quenching the coupling constant (\S\ref{sec:Morigi-qJ}), and
discuss its possible observation in a laboratory experiment (\S
\ref{sec:Morigi-exp}). Section \ref{sec:cooling} is devoted to
demonstrating that an iterative application of the previously considered
quenching protocols can cool the systems. Finally, conclusions are
drawn and open issues are discussed in Sec.\ \ref{sec:conclusions}.

\section{Temperature inversion in the Hamiltonian mean-field (HMF) model}
\label{sec:HMF}
\begin{figure}
\centering
\includegraphics[width=8cm]{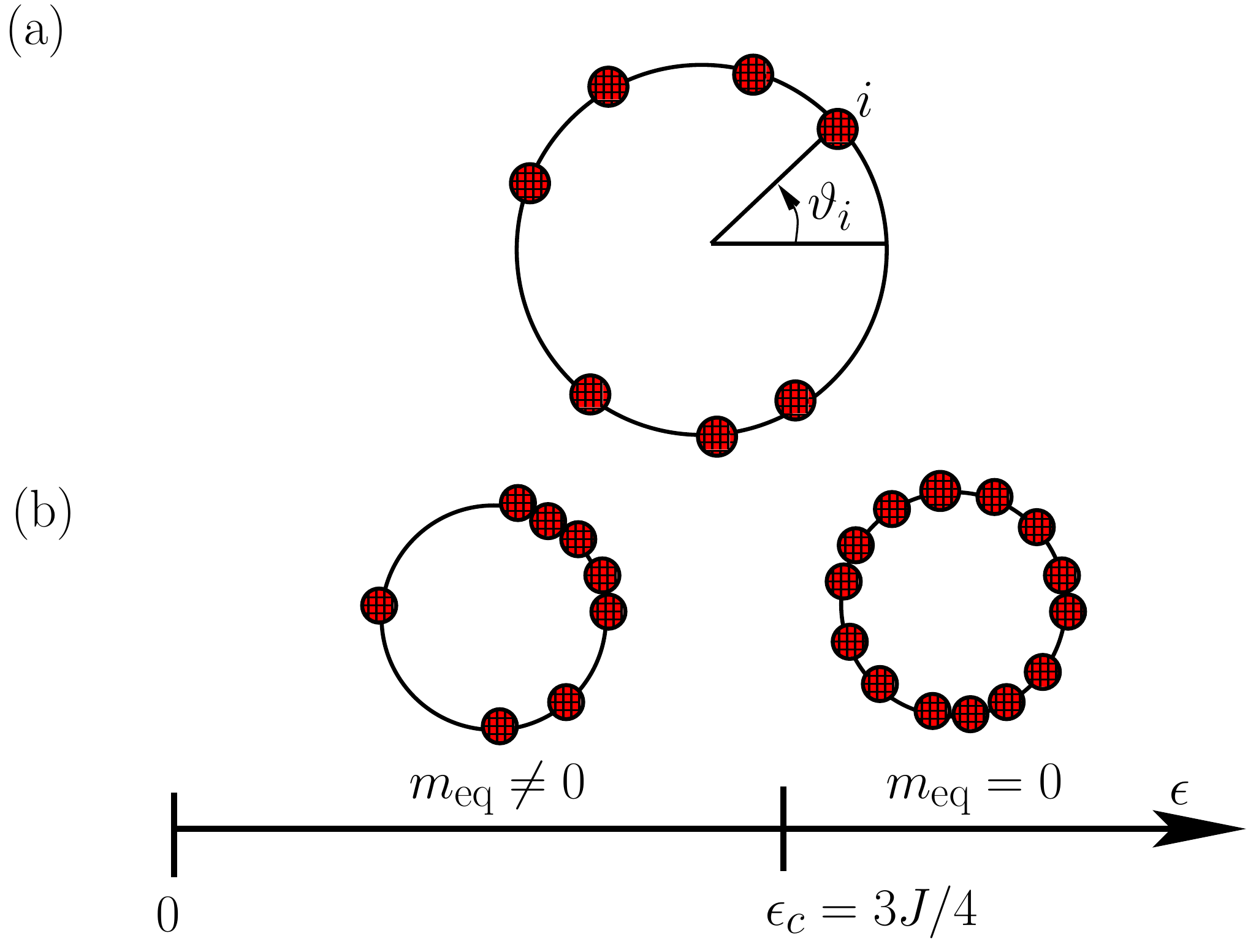}
\caption{(a) Schematic diagram of the Hamiltonian mean-field (HMF) model;
the $i$-th particle is characterized by the angular coordinate $\theta_i$ and conjugate
momentum $p_i$. (b) Equilibrium phase diagram of the F-HMF model in the
absence of the external field $h$: As a function of the energy density $\varepsilon$, a continuous
transition between a magnetized ($m_{\rm eq} \ne 0$) and a
non-magnetized ($m_{\rm eq}=0$)
state occurs at the critical value $\varepsilon_c=3J/4$.}
\label{fig:HMF-setup}
\end{figure}

The prototypical model of long-range interacting systems that we shall consider in
this paper is the so-called Hamiltonian mean field (HMF)
model \cite{Antoni:1995}. The
model comprises a system of $N$ globally-coupled point particles of unit
mass moving on a circle. The Hamiltonian of the system, in presence of
an external field of strength $h$, is given by
\be
H=\sum_{i=1}^N\frac{p_i^2}{2}+\frac{J}{2N}\sum_{i,j=1}^N
[1-\cos(\theta_i-\theta_j)]-h\sum_{i=1}^N\cos \theta_i.
\label{eq:HMF-H}
\ee
Here, $\theta_i\in [-\pi,\pi]$ is the angular coordinate of the $i$-th
particle $(i = 1,2,\ldots,N)$ on the circle, while $p_i$ is the conjugated
momentum, see Fig.\ \ref{fig:HMF-setup}(a). The coupling constant $J$ can be either positive or negative,
defining respectively the ferromagnetic (F) and the antiferromagnetic
(AF) version of the model. The reason for the use of terms borrowed from the physics of magnetic systems is that the HMF model can also be seen as a system of planar ($XY$) spins with mean-field couplings (that is, defined on a complete graph with links of the same strength). For $h = 0$, therefore, the Hamiltonian (\ref{eq:HMF-H}) is $O(2)$-invariant. Ferromagnetic (respectively, antiferromagnetic) interactions in the magnetic picture correspond to attractive (respectively, repulsive) interactions in the particle interpretation. 
The time evolution of the system is governed by the Hamilton equations derived from the Hamiltonian (\ref{eq:HMF-H}):
\be
\left\{
\begin{array}{ccl}
{\displaystyle \frac{{\rm d}\theta_i}{{\rm d}t}}& = & p_i\,,\\
& & \\ 
{\displaystyle \frac{{\rm d}p_i}{{\rm d}t}}& = & J(m_y \cos \theta_i-m_x \sin \theta_i)-h\sin \theta_i\,, 
\end{array}
\right.
\label{eq:EOM} 
\ee
where $(m_x ,m_y)\equiv (1/N)\sum_{i=1}^N(\cos \theta_i,\sin
\theta_i)$ are the components of the magnetization vector
${\bf m}\equiv (m_x, m_y)$, whose magnitude $m\equiv\sqrt{m_x^2+m_y^2}$
measures the amount of clustering of particles on the circle. A uniform
(respectively, non-uniform) distribution of particles on the circle
implies the value $m=0$ (respectively, $m\ne0$).

Let us denote by $f(\theta,p,t)$ the time-dependent single-particle
distribution function, which is the probability density to have a particle with
angular coordinate $\theta$ and momentum $p$ at time $t$. In thermal equilibrium\footnote{Despite long-range interactions implying a non-additive Hamiltonian, in the HMF model in the thermodynamic limit, canonical and microcanonical ensembles are equivalent for any temperature and energy density, so that one can speak of thermal equilibrium without specifying if it is a canonical or a microcanonical equilibrium \cite{Campa:2009,Campa:2014}.} at temperature $T$, the distribution has the form \cite{Campa:2014}
\be
f_{\rm eq}(\theta,p) \propto \exp\left\{-\frac{1}{T}\left[\frac{p^2}{2}-J\left(m_x^{\rm eq}\cos \theta+m_y^{\rm eq}\sin \theta\right)-h\cos \theta\right]\right\},
\label{eq:HMF-Gaussian-solution}
\ee 
where the equilibrium magnetization components are obtained
self-consistently as $(m_x^{\rm eq}, m_y^{\rm eq})\equiv \int {\rm
d}\theta {\rm d}p \,(\cos \theta, \sin \theta)f_{\rm eq}(\theta,p)$. In Eq.\ (\ref{eq:HMF-Gaussian-solution}), we have set the Boltzmann constant to unity, as we shall do throughout the paper. In the special case $m_y^{\rm eq}= 0$, that is, for distribution of particles symmetric around $\theta = 0$ that we shall focus on in the following, the self-consistent relation for the magnetization may be written as  \cite{Campa:2009,Campa:2014} 
\be
m_{\rm eq}= \frac{I_1\left[\beta (J m_{\rm eq}+h)\right]}{I_0\left[\beta (J m_{\rm eq}+h)\right]}~,
\ee
where $m_{\rm eq}\equiv m_x^{\rm eq},\beta = T^{-1}$, and $I_n(x)$
denotes the modified Bessel function of order $n$. For $h\ne0$, the
system in thermal equilibrium is characterized by an inhomogeneous
distribution of particles on the circle, with clustering around
$\theta=0$. In the AF case, a nonzero external field is necessary to
have an inhomogeneous thermal state, while in the F case, a clustered equilibrium state is attained even with $h=0$, via a spontaneous breaking of the $O(2)$ symmetry by the attractive interaction between the particles, provided the total energy density is smaller than a critical value
$\varepsilon_c \equiv 3J/4$, see Fig.\ \ref{fig:HMF-setup}(b); the corresponding critical temperature is $T_c\equiv J/2$ \cite{Campa:2009,Campa:2014}. The F-HMF system with $h=0$ in fact shows a continuous phase transition as a function of temperature, with $m_{\rm eq}$ decreasing continuously from unity at $T=0$ to zero at $T=T_c$, while remaining zero at higher temperatures \cite{Campa:2009,Campa:2014}.  

Let us note that the Hamiltonian (\ref{eq:HMF-H}) can be obtained by
taking any long-range interacting system, restricting to one spatial
dimension, expanding the potential energy in a Fourier series, and
retaining just the first Fourier mode \cite{Levin:2014,Teles:2015}. The
AF-HMF model can be seen to represent a one-component Coulomb system,
while the F-HMF is a simplified description of a self-gravitating
system. As anticipated in the Introduction, and as we shall see further in Sec.\ \ref{sec:Morigi}, the F-HMF model turns out to be very closely related to a different system, i.e., a system of atoms in an optical cavity.

\subsection{The antiferromagnetic case}
Let us now discuss how quenching one of the parameters of the Hamiltonian (\ref{eq:HMF-H}) when the system is in an inhomogeneous thermal equilibrium state produces a non-equilibrium state exhibiting temperature inversion. To start with, we consider the AF case with $h \not = 0$.

\subsubsection{Quenching the external field}
\label{sec:AF-qh}
\begin{figure}
\begin{center}
\centering \includegraphics[width=10cm]{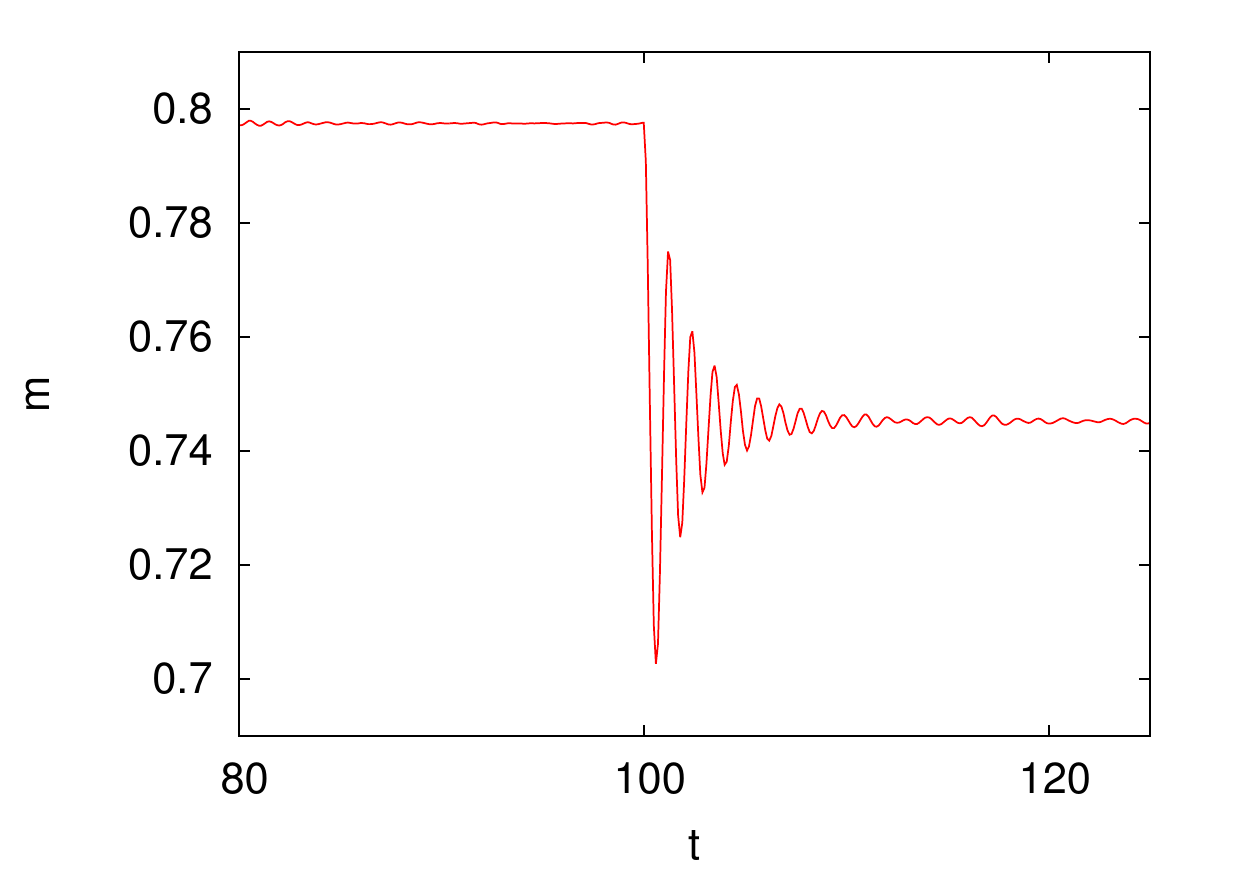}
\end{center}
\caption{AF-HMF model in presence
of an external field. Time evolution of the magnetization $m$: Starting with thermal equilibrium (\ref{eq:HMF-Gaussian-solution}) at temperature
$T=5$ and field $h=15$, the field strength is instantaneously
quenched at $t=100$ to $h=10$. The number of particles is $N=10^6$, and the coupling constant is $J=-1$.}
\label{fig:ahmf-quench-h-mag}
\end{figure}

\begin{figure}
\begin{center}
\centering \includegraphics[width=10cm]{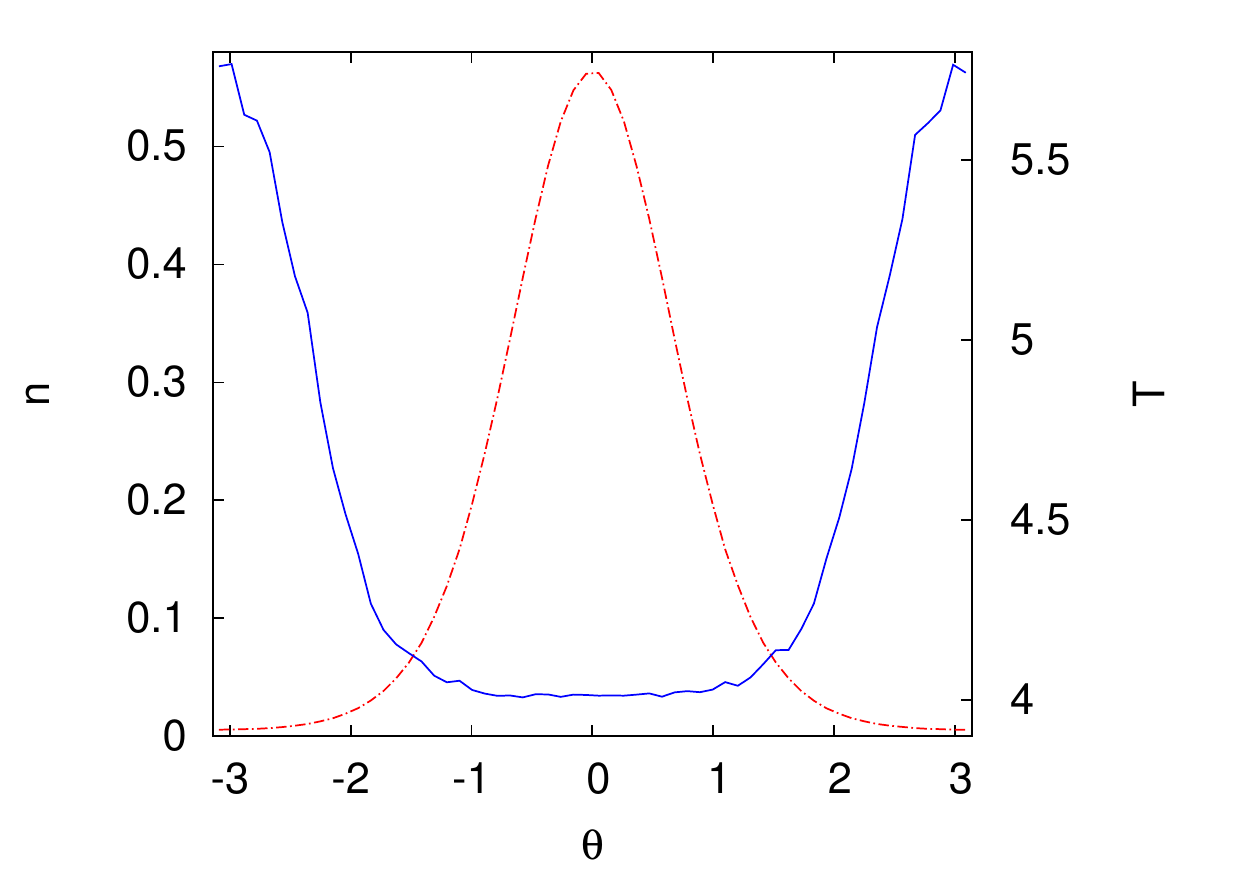}
\end{center}
\caption{Temperature inversion in the AF-HMF model after a quench of the
external field. Local density $n$ (red dot-dashed line) and local temperature $T$ (blue solid line) measured at $t = 10^3$ in the QSS obtained under the same conditions as for Fig.\
\ref{fig:ahmf-quench-h-mag}.}
\label{fig:ahmf-quench-h-temp-inv}
\end{figure}

We consider $N = 10^6$ particles and prepare the system in an
inhomogeneous equilibrium state ($m_{\rm eq}\ne 0$) at an initial value of the field $h=15$ and temperature $T=5$, by sampling independently for every particle the coordinate $\theta$ and the momentum $p$ from the distribution (\ref{eq:HMF-Gaussian-solution}). We take the coupling to be $J=-1$. In such a state, while the local density $n(\theta)$, defined as
\be
n(\theta)\equiv \int {\rm d}p\,f(\theta,p,t)\,,
\label{eq:n-defn}
\ee
is non-uniform, corresponding to a magnetization $m_{\rm eq}\approx 0.797$, the local temperature, defined as
\be
T(\theta)\equiv \frac{\int {\rm d}p\,p^2 f(\theta,p,t)}{n(\theta)}\,,
\label{eq:Ttheta-defn}
\ee
is nevertheless uniform as a function of $\theta$. We study the time
evolution of the system starting from this equilibrium state by
performing a MD simulation that involves integrating\footnote{In all the
MD simulations reported in this paper, we used a fourth-order symplectic
algorithm \cite{McLachlan:1992}, with time step $\delta t=0.1$. This
ensured energy conservation up to a relative fluctuation of $10^{-7}$.}
the equations of motion (\ref{eq:EOM}). At $t = 100$, we instantaneously
quench the field to $h=10$. Soon after the quench, the magnetization of
the system starts oscillating; the oscillations damp out in time, and
eventually the system relaxes to a QSS with $m_{\rm QSS} < m_{\rm eq}$
at the new value of the field. The latter fact is expected since the value of $h$ is smaller than in the initial state. The time evolution of the magnetization for a time window around the time of the quench is shown in Fig.\ \ref{fig:ahmf-quench-h-mag}.
For times larger than those shown in Fig.\ \ref{fig:ahmf-quench-h-mag},
the magnetization stays essentially constant with extremely small
fluctuations, signaling that the system has relaxed to a QSS. The
density and temperature profiles as a function of $\theta$ in the QSS
are plotted in Fig.\ \ref{fig:ahmf-quench-h-temp-inv}. It is apparent
that $n(\theta)$ and $T(\theta)$ are anticorrelated. The system exhibits a clear temperature inversion: denser regions are colder than sparser ones. We note that the average temperature 
\be
\langle T \rangle \equiv \int  {\rm d}\theta\,T(\theta)n(\theta)
\ee
has the value $\langle T \rangle \approx 4$ in
 the QSS, lower than the initial value $T = 5$. We shall come back to this point in \S\ref{sec:cooling}.

\begin{figure}
\centering \includegraphics[width=18cm]{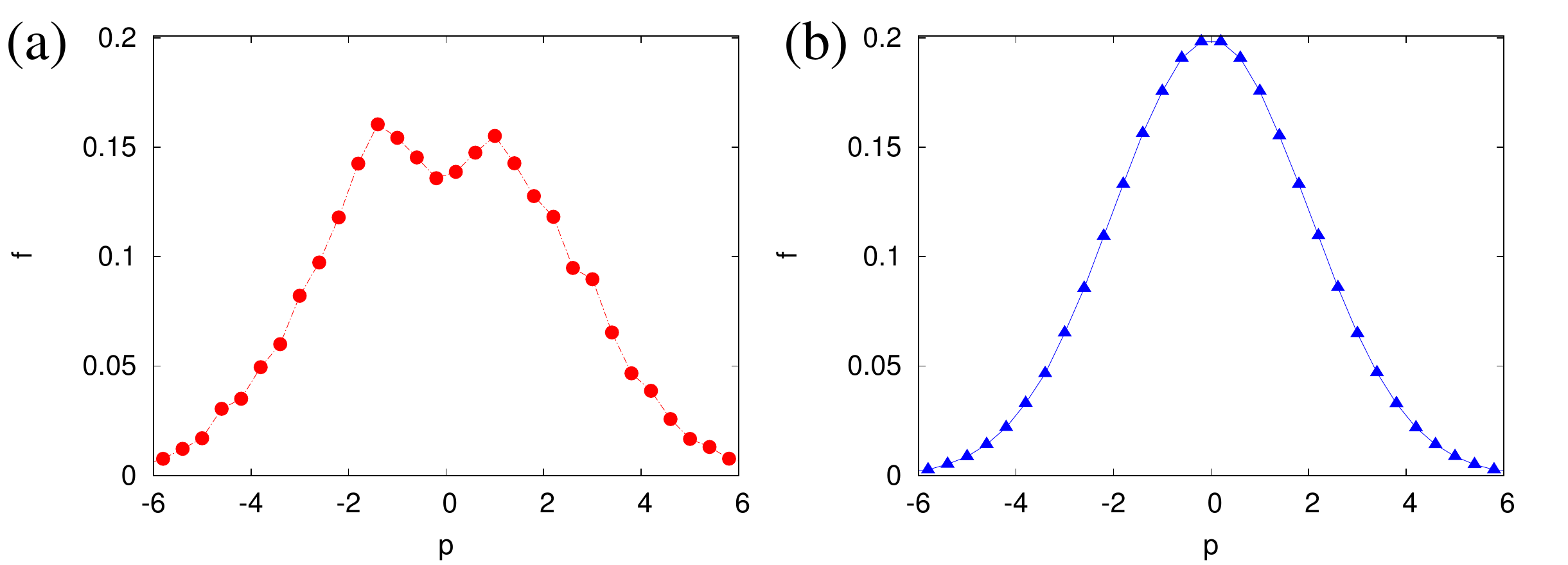}
\caption{Momentum distribution of the AF-HMF model measured at $t = 10^3$ in the QSS obtained under the same conditions as for Fig.\
\ref{fig:ahmf-quench-h-mag}. (a) The distribution at $\theta=\pi$ at
which the density is minimum. (b) The distribution at $\theta=0$ at which the density is maximum.}
\label{fig:ahmf-quench-h-f(p)}
\end{figure}

Apart from the reduction of the average temperature, these results are
qualitatively very similar to those obtained in \cite{Teles:2015} for a
ferromagnetic HMF model. In that work, however, the simulation protocol was
different, because the system was initiated in equilibrium with $h = 0$
and then an external field was applied for a very short time. In Ref.\
\cite{Teles:2015}, a physical picture was put forward to explain the
emergence of temperature inversion, and we argue that it applies also in
the present case (and, as we shall discuss below, to all the cases
discussed in the present paper). Without entering into details, let us
recall the main points of this explanation. After the quench (or the
perturbation), a collective oscillation (i.e., a wave) sets in, as
witnessed by the time evolution of the magnetization (Fig.\
\ref{fig:ahmf-quench-h-mag}). The oscillation damps out in time, but the
system is conservative with no dissipative mechanism being present.
Hence, the energy lost by the wave must be acquired by some of the
particles. This may happen via Landau damping\footnote{The theory of
Landau damping is completely understood for homogeneous states and small
perturbations \cite{plasma,Villani:2011}; its extension to inhomogeneous
states still has many open issues \cite{Barre:2010,Barre:2011}. However,
the basic physical mechanism works in any situation where particles
interact with a collective excitation in the system.}, an ubiquitous
phenomenon in systems with long-range interactions: particles whose
velocity is not too far from the phase velocity $v_{\rm ph}$ of the wave
interact strongly with the wave itself. The net result is that the
nearly resonant particles gain kinetic energy from the wave, so that the
momentum distribution $f(p)\equiv \int {\rm d}\theta f(\theta,p)$ develops small peaks or shoulders and
becomes different from a thermal distribution. In particular, close to
the resonances, there will be suprathermal regions, i.e., regions where
$f(p)$ is larger than the initial thermal distribution. Now, the
mechanism of velocity filtration comes into play as follows.  The
density $n(\theta)$ is maximum at $\theta = 0$ and decreases for larger
and smaller $\theta$'s, reaching its minimum around $\theta = \pm \pi$.
The net effect of the interaction between the particles and the external
field is to create an effective force field pushing the particles
towards $\theta = 0$, so that the particles have to climb the potential
energy well to reach the sparser regions of the system; particles with
larger kinetic energies will do that more easily than ``colder''
particles, and any suprathermal region present in the velocity
distribution will be magnified in regions where the particle density is
lower. This is precisely what happens in our case, as shown in Fig.\
\ref{fig:ahmf-quench-h-f(p)}, where the velocity distribution function
measured at a position where the particle density $n$ is smaller,
$f(\theta=\pi,p)$, see panel (a), is plotted together with the same
function measured at the position of maximum density, $f(\theta=0,p)$,
see panel (b). Magnified suprathermal regions in $f(\theta=\pi,p)$ are apparent, and are responsible for the fact that the variance of the velocity distribution is here larger than in the denser regions of the system.

We performed other numerical experiments with the same quench protocol
by varying the initial temperature and the difference between the
initial and final values of the field $h$, and obtained qualitatively similar results. No fine-tuning of the parameters is necessary to observe temperature inversion. We repeated some of the numerical experiments with $N = 10^7$ particles, observing no differences with respect to the $N = 10^6$ case.

In principle, one can quench also the coupling constant $J$, while
keeping the field $h$ fixed. However, as stated before, as far as
laboratory systems are concerned, the AF-HMF system can be seen as a toy
model of a system of equal electric charges in a confining field, i.e.,
ions in a trap. In
trapped ions, quenching $h$ corresponds to quenching the strength of the
confining trap, which in principle is an experimentally feasible
protocol. Conversely, quenching $J$ is equivalent to quenching the charge
of the ions, which does not seem very physical, so that we do not study this kind of quench protocol for the AF case.
\subsection{The ferromagnetic case}
Let us now turn to the ferromagnetic (F-HMF) case, that is, $J > 0$. Without loss of generality, we take the coupling to be $J=1$.

\subsubsection{Quenching the external field}
\label{sec:F-qh}
\begin{figure}
\begin{center}
\centering \includegraphics[width=10cm]{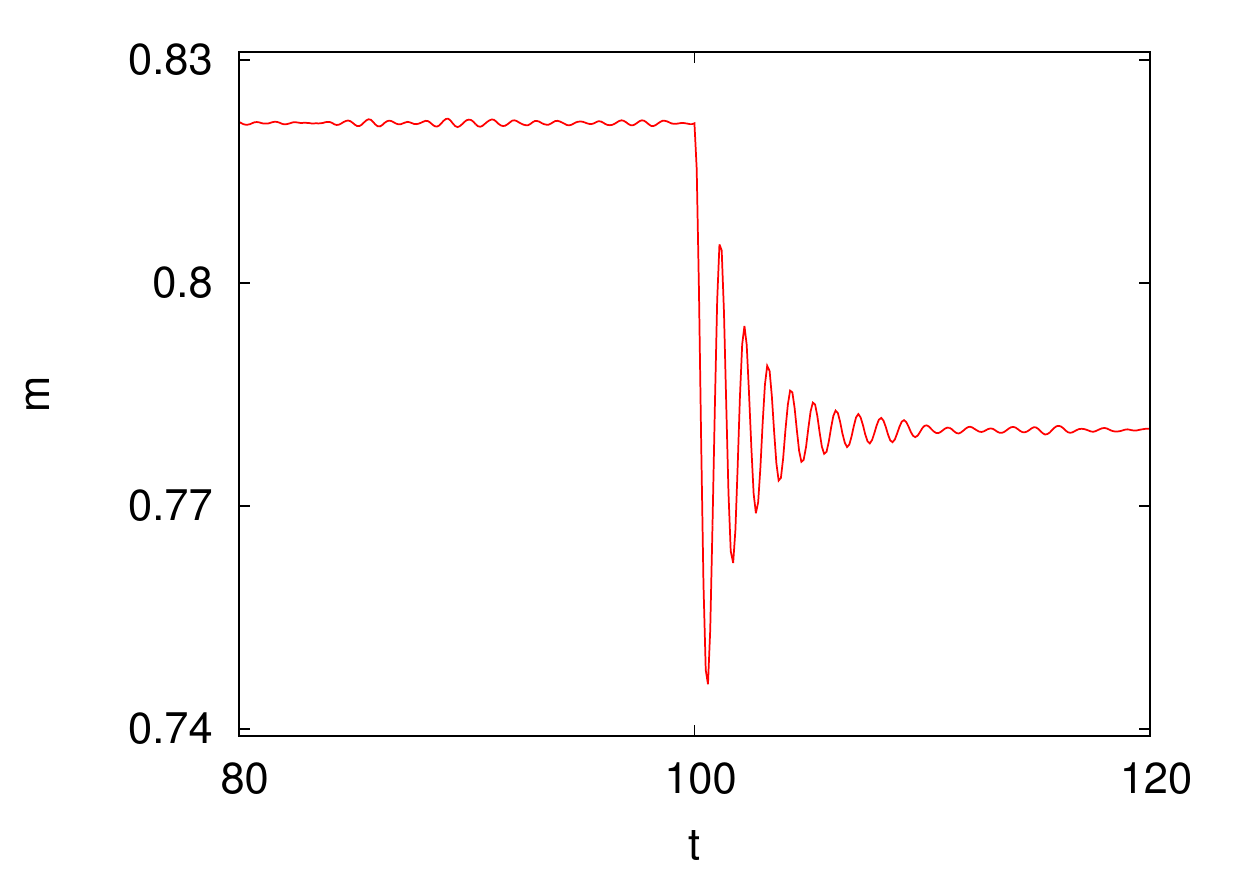}
\end{center}
\caption{F-HMF model in presence
of an external field. Time evolution of the magnetization $m$: Starting with thermal equilibrium (\ref{eq:HMF-Gaussian-solution}) at temperature
$T=5$ and field $h=15$, the field is instantaneously
quenched at $t=100$ to $h=10$. The number of particles is $N=10^6$, and the coupling constant is $J=1$.}
\label{fig:hmf-quench-h-mag}
\end{figure}

\begin{figure}
\begin{center}
\centering \includegraphics[width=10cm]{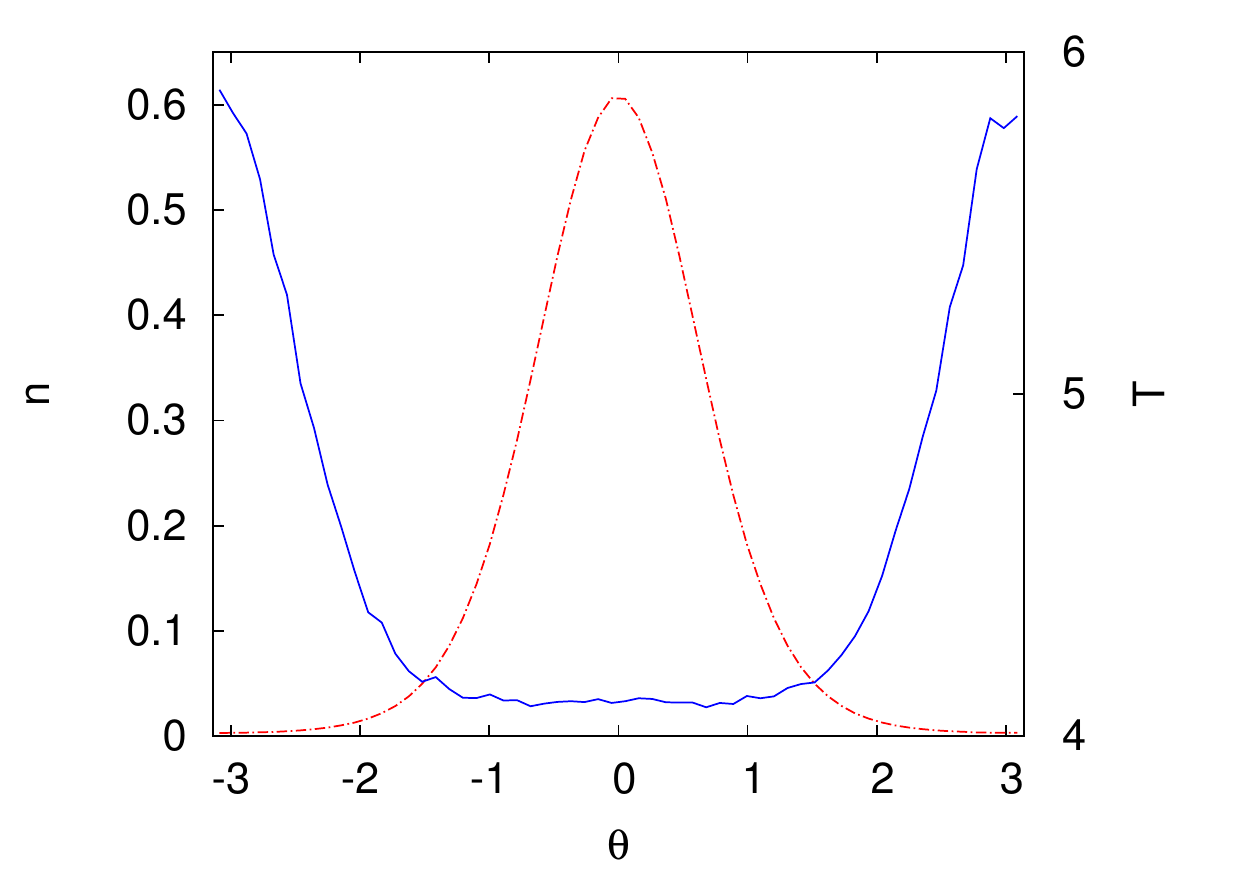}
\end{center}
\caption{Temperature inversion in the F-HMF model after a quench of the external field. 
Local density $n$ (red dot-dashed line) and local temperature $T$ (blue
solid line) measured at $t = 10^3$ in the QSS obtained under the same conditions as for Fig.\
\ref{fig:hmf-quench-h-mag}.}
\label{fig:hmf-quench-h-temp-inv}
\end{figure}

We repeat for the F-HMF system the numerical experiment we performed
with the AF system. We prepare the system of $N = 10^6$ particles in
inhomogeneous thermal equilibrium at temperature $T=5$ and with a field
$h=15$, now corresponding to $m_{\rm eq}\approx 0.821$, and evolve the
system up to $t = 100$, when we quench the external field to $h = 10$.
As in the AF case, the magnetization starts oscillating (Fig.\
\ref{fig:hmf-quench-h-mag}), the oscillation damps out in time, and the system settles in a QSS with temperature inversion (Fig.\ \ref{fig:hmf-quench-h-temp-inv}). 

\begin{figure}
\centering \includegraphics[width=18cm]{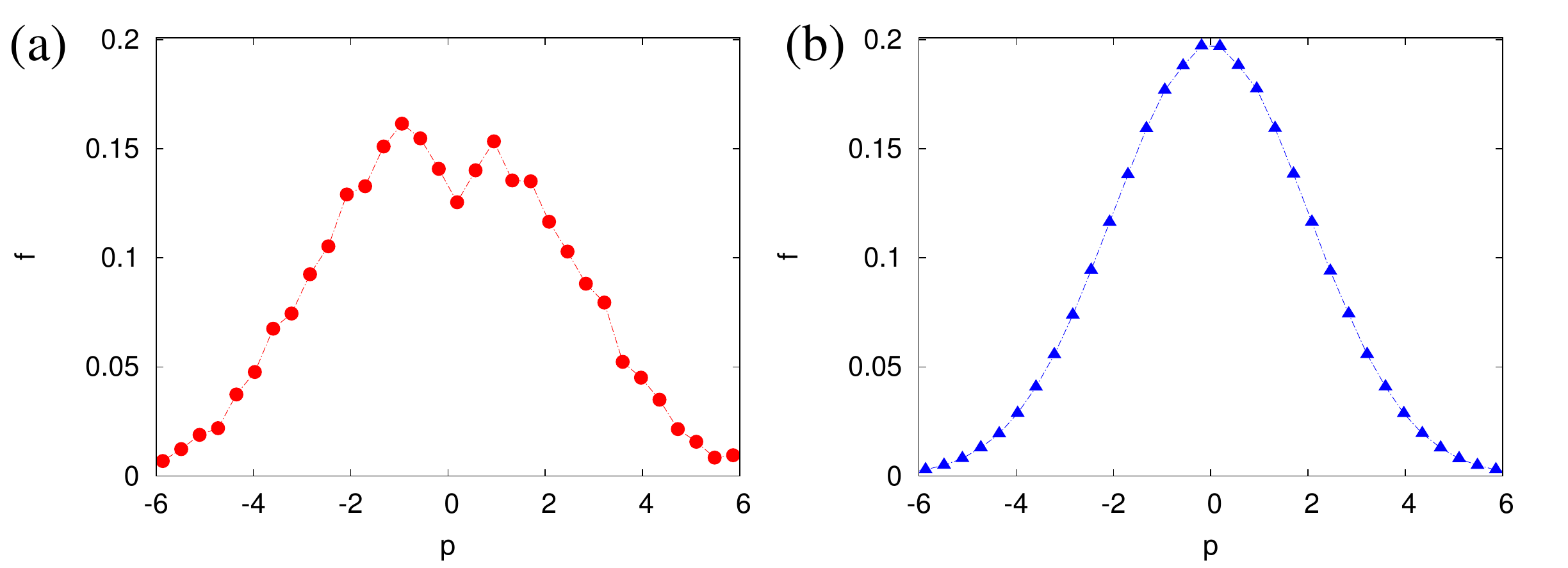}
\caption{Momentum distribution of the F-HMF model measured at $t = 10^3$ in the QSS obtained under the same conditions as for Fig.\
\ref{fig:hmf-quench-h-mag}. (a) The distribution at $\theta=\pi$ at
which the density is minimum. (b) The distribution at $\theta=0$ at which the density is maximum.}
\label{fig:hmf-quench-h-f(p)}
\end{figure}

Figures \ref{fig:hmf-quench-h-mag} and \ref{fig:hmf-quench-h-temp-inv}
are very similar to the corresponding ones for the AF system, that is,
Figs.\ \ref{fig:ahmf-quench-h-mag} and \ref{fig:ahmf-quench-h-temp-inv},
respectively. The same is true for the momentum distributions, shown in
Fig.\ \ref{fig:hmf-quench-h-f(p)}, which is strikingly similar to Fig.\ \ref{fig:ahmf-quench-h-f(p)}. 
Hence, also in this case, the physical picture based on wave-particle
interaction and velocity filtration applies. Consistent with our
physical picture, the sign of the interactions is irrelevant for field
quenches, as long as the field is sufficiently strong to produce the
dominant effects of shaping the density profile $n(\theta)$ and
providing the effective potential well that filters the velocities of
the particles, thereby producing temperature inversion. Again, we
repeated our numerical experiments for different values of the fields,
for different initial temperatures and for systems with $N = 10^7$,
observing no qualitative differences. We note that similar to the AF
case, the average temperature in the QSS for the F case is smaller than
the initial temperature, and one has $\langle T \rangle \approx 4$.

\subsubsection{Quenching the coupling constant}
\label{sec:F-qJ}
\begin{figure}[!h]
\begin{center}
\centering \includegraphics[width=10cm]{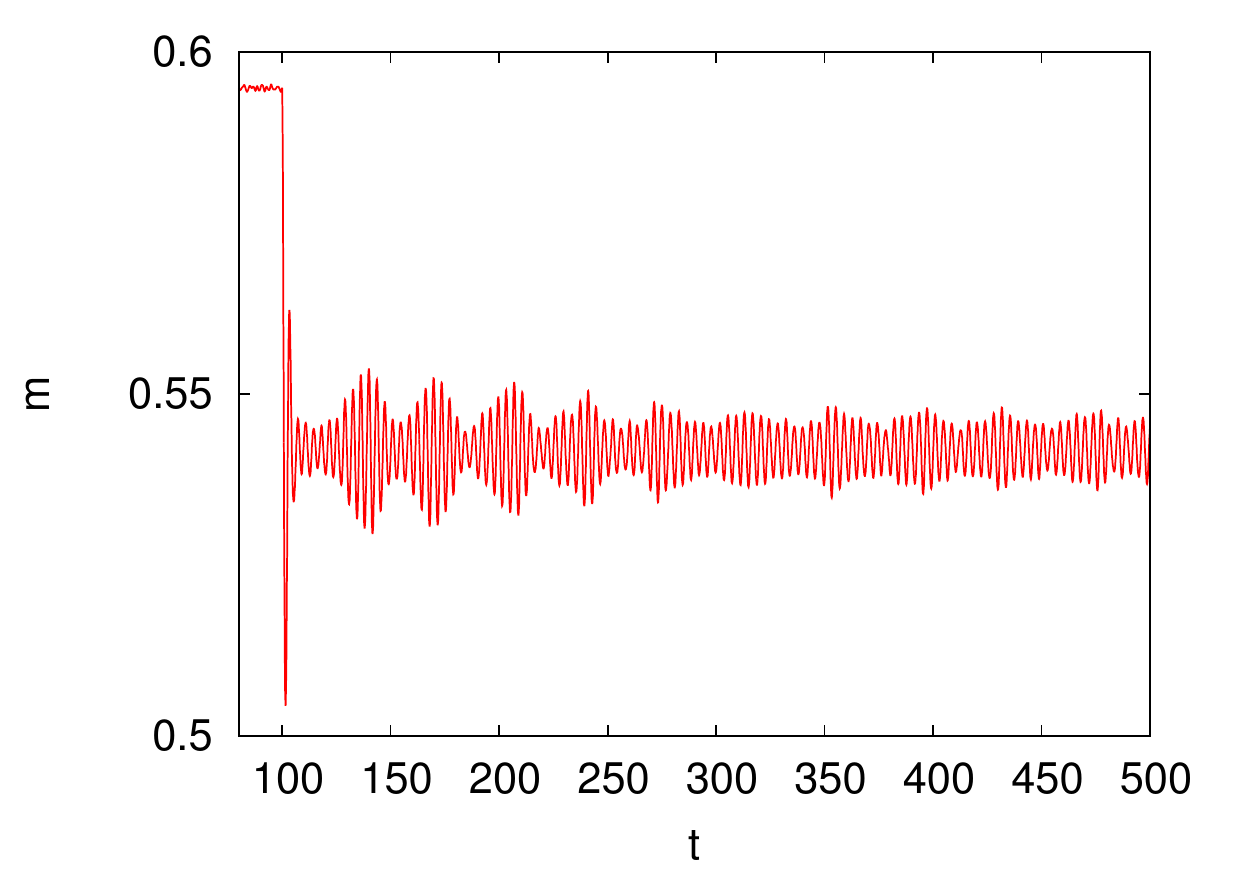}
\end{center}
\caption{F-HMF model without an external field. Time evolution of the
magnetization $m$: Starting with thermal equilibrium
(\ref{eq:HMF-Gaussian-solution}) at temperature $T=2$ and coupling
constant $J=5$, the coupling is instantaneously
quenched at $t=100$ to $J=4$. The number of particles is $N=10^6$.}
\label{fig:hmf-quench-J-mag}
\end{figure}

\begin{figure}[!h]
\begin{center}
\centering \includegraphics[width=10cm]{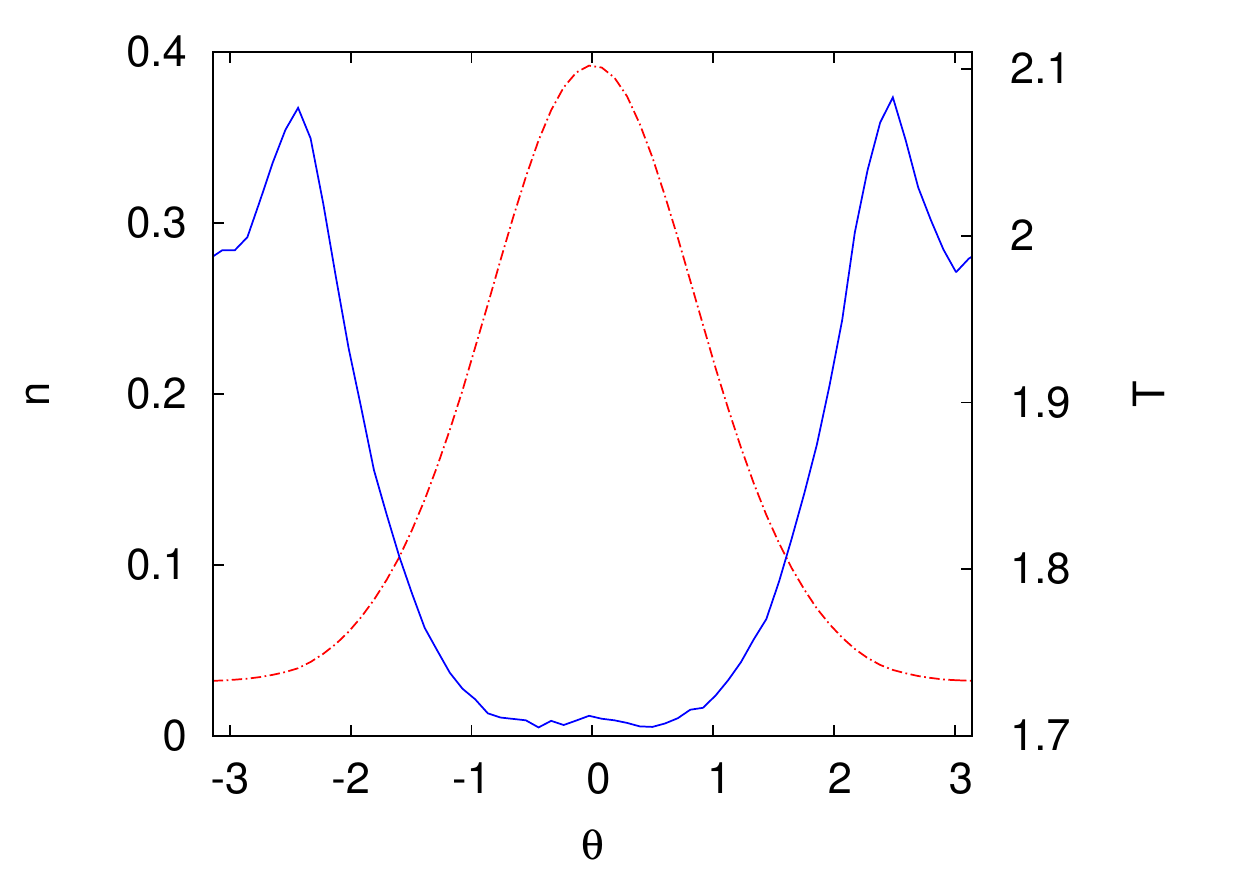}
\end{center}
\caption{Temperature inversion in the F-HMF model after a quench of the
coupling. Local density $n$ (red dot-dashed line) and local temperature
$T$ (blue solid line) measured at $t = 10^3$ in the QSS obtained under
the same conditions as for Fig.\ \ref{fig:hmf-quench-J-mag}.}
\label{fig:hmf-quench-J-temp-inv}
\end{figure}

\begin{figure}[!h]
\centering \includegraphics[width=18cm]{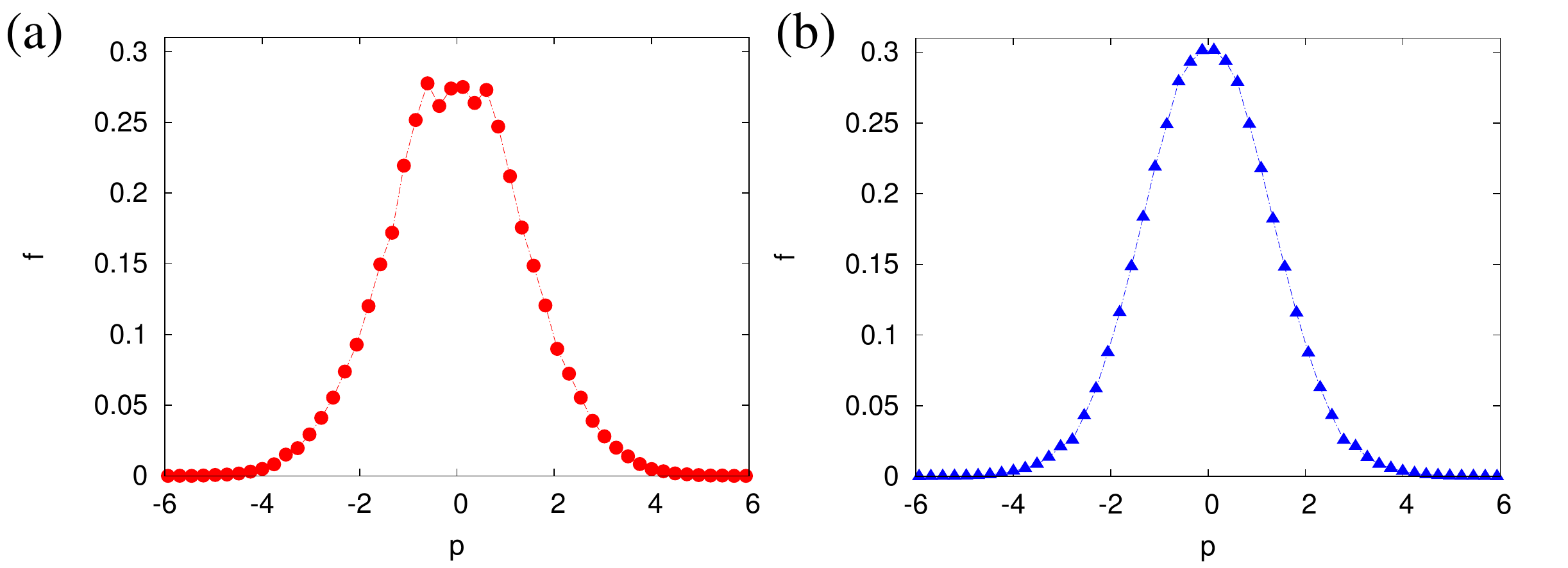}
\caption{Momentum distribution of the F-HMF model measured at $t = 10^3$ in the QSS obtained under the same conditions as for Fig.\
\ref{fig:hmf-quench-J-mag}. (a) The distribution at $\theta=\pi$ at
which the density is minimum. (b) The distribution at $\theta=0$ at which the density is maximum.}
\label{fig:hmf-quench-J-f(p)}
\end{figure}

As we have previously discussed, the F-HMF model admits magnetized (collapsed) thermal equilibrium states also for $h = 0$, provided $T < J/2$. Moreover, as we shall see in Sec.\ \ref{sec:Morigi}, the F-HMF model also admits an interpretation in terms of an atomic system, where the coupling constant $J$ is tunable in an experiment. We thus turn to analyze what happens if we prepare an F-HMF system in thermal equilibrium with nonzero magnetization and $h = 0$, and then quench the coupling constant.
We prepare the system of $N = 10^6$ particles in inhomogeneous thermal
equilibrium at temperature $T=2$ and coupling $J=5$, by sampling independently
for every particle the coordinate $\theta$ and the momentum $p$ from the
distribution (\ref{eq:HMF-Gaussian-solution}). As before, we evolve the
system until $t = 100$, when we instantaneously quench the coupling to
$J=4$. As shown in Figs.\ \ref{fig:hmf-quench-J-mag},
\ref{fig:hmf-quench-J-temp-inv} and \ref{fig:hmf-quench-J-f(p)}, the
behavior of the system is similar to the case when the external field
$h$ was quenched; although the effect is quantitatively less dramatic,
still there is a clear temperature inversion in the QSS, where the
average temperature has the value $\langle T \rangle \approx 1.75$,
lower than the initial value of $2$. Note that in contrast to the case when we quenched the strength of the field, here the damping of the oscillations in $m(t)$ with time is not quite complete (see Fig.\ \ref{fig:hmf-quench-J-mag}); indeed, oscillations are sustained also at long times, though the amplitude of oscillation is small ($\sim 0.005$ at $t = 500$, compared to the amplitude $\sim 0.1$ subsequent to the quench).
The physical picture based on wave-particle interactions and velocity
filtration holds also in this case. We performed other numerical
experiments with larger $N$'s and different values of $J$ and, as
before, we can conclude that no fine tuning is needed to produce
temperature inversion by means of a quench protocol, acting on either $h$ or on $J$.

Let us now discuss how the F-HMF model is related to a system of atoms interacting with a standing electromagnetic wave in an optical cavity, and how a quench of the coupling constant could be performed in a controlled way in a laboratory experiment. 

\section{Temperature inversion in a system of atoms in an optical cavity}
\label{sec:Morigi}
\begin{figure}[!h]
\centering
\includegraphics[width=10cm]{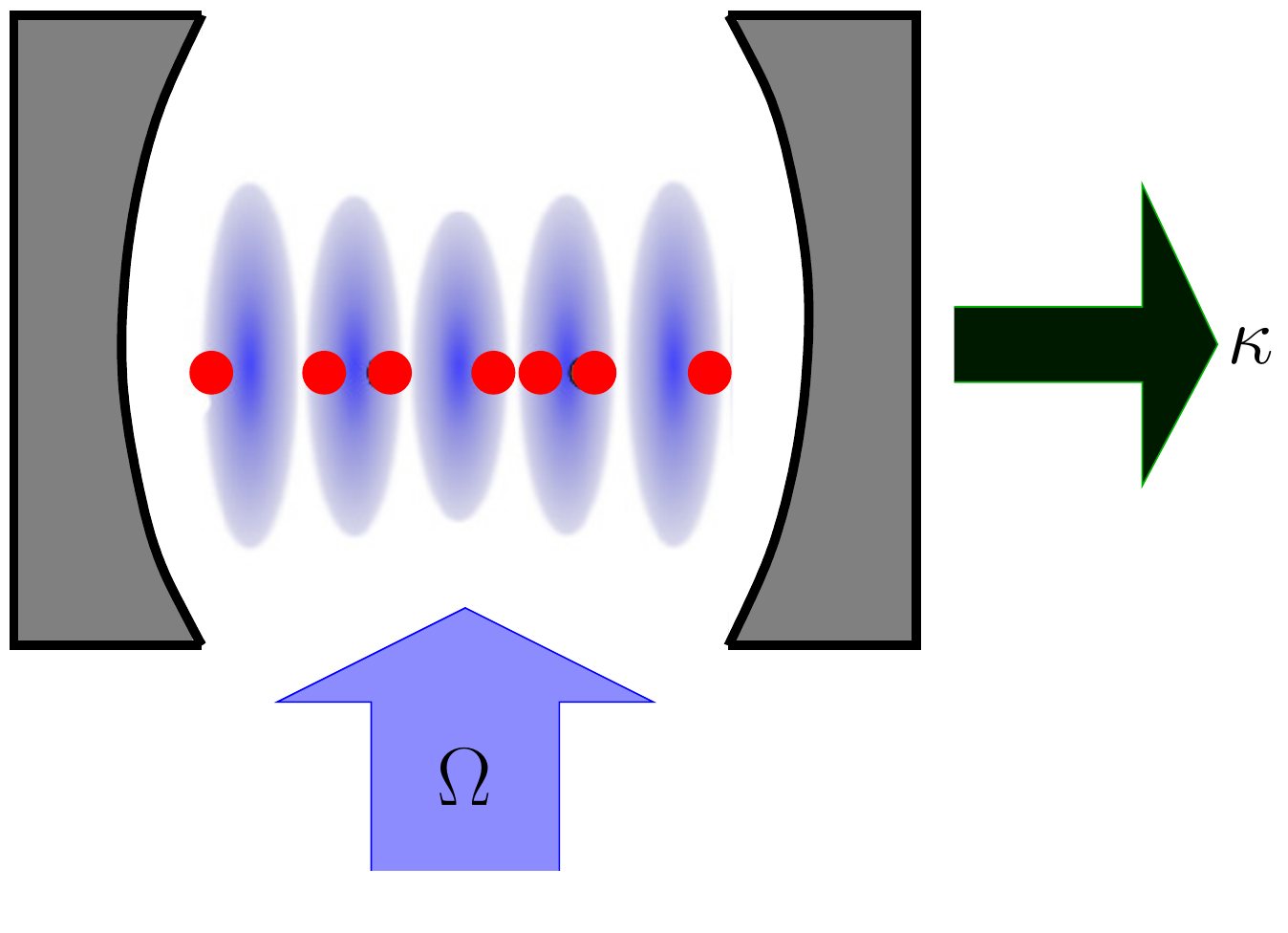}
\caption{Atoms interacting with a single-mode standing electromagnetic wave in a cavity
of linewidth $\kappa$, and being driven by a transverse laser with
intensity $\Omega^2$.}
\label{fig:morigi-setup}
\end{figure}

Atoms interacting with a single-mode standing electromagnetic wave due
to light trapped in a high-finesse optical cavity are subject to an
interparticle interaction that is long-ranged owing to multiple
coherent scattering of photons by the atoms into the wave mode
\cite{Schutz:2014,Schutz:2015,Jager:2016}. The system serves as a unique
platform to study long-range interactions, and in particular mean-field interactions,  under tunable experimental
conditions. The typical setup is sketched in Fig.\ \ref{fig:morigi-setup}, which also
shows optical pumping by a transverse laser of intensity $\Omega^2$ to
counter the inevitable cavity losses quantified by the cavity linewidth
$\kappa$ (the lifetime of a photon in the cavity being $\kappa^{-1}$).
We follow Ref. \cite{Schutz:2014,Schutz:2015,Jager:2016}, and refer
the reader to these works and to references therein for all the
technical details
that we skip here. Let us then consider a system of $N$ identical atoms
of mass $m$. As far as the interactions with the electromagnetic
field is concerned, each atom may be regarded as a two-level system, where the
transition frequency between the two levels is $\omega_0$. If the atoms
are confined in one dimension along the cavity axis (taken to be the
$x$-axis), and $k$ is the wavenumber of the standing wave, the sum of the
electric field amplitudes coherently scattered by the atoms at time $t$
depends on their instantaneous positions $x_1,\ldots,x_N$, and is proportional to the quantity 
\be
\Theta\equiv\frac{1}{N}\sum_{j=1}^N\cos(kx_j)\,,
\label{eq:Theta-defn}
\ee
so that the cavity electric field at time $t$ is $E(t)\propto
\Theta\sqrt{N\overline{n}}$ \cite{Schutz:2014}. Here, $\overline{n}$ is
the maximum intracavity-photon number per atom, given by
$\overline{n}\equiv N\Omega^2\alpha^2/(\kappa^2+\Delta_c^2)$, with
$\alpha \equiv g/\Delta_a$ being the ratio between the cavity vacuum
Rabi frequency and the detuning $\Delta_a\equiv \omega_L-\omega_0$
between the laser and the atomic transition frequency, and $\Delta_c\equiv
\omega_L-\omega_c$ being the detuning between the laser and the
cavity-mode frequency. It is important to note that $\overline{n}$ is tunable by
controlling either the strength of the external laser pump or the detuning $\Delta_c$. The quantity $\Theta$ characterizes the amount of spatial ordering of atoms within
the cavity mode, with $\Theta=0$ corresponding to atoms being uniformly
distributed and the resulting vanishing of the cavity field, and
$|\Theta| \ne 0$ implying spatial ordering. In particular, $|\Theta|=1$
corresponds to perfect ordering of atoms to form Bragg gratings and the
resulting scattering of photons in phase, thereby maximizing the cavity
field, which in turn traps the atoms by means of mechanical forces. Note
that the wave number $k$ is related to the linear dimension $L$ of the cavity
through $k=2\pi/\lambda$ and $L=q\lambda$, where $\lambda$ is the
wavelength of the standing wave, and $q \in \mathbb{N}$.

The dynamics of the system may be conveniently studied by analyzing the
time evolution of the $N$-atom phase space distribution
$f_N(x_1,\ldots,x_N, p_1,\ldots,p_N,t)$ at time $t$, with $p_j$'s
denoting the momenta conjugate to the positions $x_j$. Treating the
cavity field quantum mechanically, and regarding the atoms as
classically polarizable particles with semiclassical center-of-mass
dynamics, it was shown that the distribution $f_N$ evolves in time according to the Fokker-Planck equation (FPE) \cite{Schutz:2014,Schutz:2015}
\be
\partial_t f_N+\{f_N,H\}=-\overline{n}\Gamma
\mathcal{L} f_N\,.
\label{eq:FPE}
\ee
Here, the operator $\mathcal{L}$ 
describes the dissipative processes (damping and diffusion) that
explicitly depend on the positions and momenta of the atoms\footnote{We
omit the explicit expression of $\mathcal{L}$ because it is not relevant
in the context of the present paper; It may be found in Ref. \cite{Schutz:2014,Schutz:2015}.}, and $\Gamma \equiv 8\omega_r \kappa \Delta_c/(\Delta_c^2+\kappa^2)$, where
$\omega_r \equiv \hbar k^2/(2m)$ is the
recoil frequency due to collision between an atom and a photon, and
$\hbar$ is the
reduced Planck constant.
The Hamiltonian $H$ is given by \cite{Schutz:2014,Schutz:2015}
\be
H = \sum_{j=1}^{N}\frac{p_{j}^{2}}{2m}+N\hbar \Delta_c \overline{n}\,\Theta^2\,.
\label{eq:H}
\ee
This semiclassical limit is valid under the condition of $\kappa$ being
larger than $\omega_r$, 
and the Hamiltonian $H$
describes the conservative dynamical evolution of $f_N$ in the limit of
vanishing cavity losses, or for times sufficiently small such that dissipative effects are negligible. The Hamiltonian (\ref{eq:H}) contains the photon-mediated long-ranged
(mean-field) interaction between the atoms encoded in the quantity $\Theta$.
Note that the interaction is attractive (respectively, repulsive) when $\Delta_c$ is negative (respectively, positive). On time scales sufficiently long so that the effect of the right-hand-side of Eq.\ (\ref{eq:FPE}) is non-negligible, the mean-field description of the dynamics is no longer valid \cite{Schutz:2014,Jager:2016}.

\subsection{The dissipationless limit and the connection with the HMF model}
\label{sec:dissipless}

Let us now consider the case of effective attractive interactions
between the atoms and the cavity field ($\Delta_c < 0$), and study the
dynamics of the system in the limit in which the effect of the
dissipation can be neglected, that is, for sufficiently small times. In
this limit, the dynamics of the $N$ atoms is conservative and governed by the Hamiltonian $H$ given by Eq.\ (\ref{eq:H}). The positions $x_j$ of the atoms enter the Hamiltonian only as $kx_j$, so that we may define the phase variables
\be
\theta_j = k x_j = 2\pi \frac{x_j}{\lambda}~,
\ee
for $j = 1,\ldots,N$. Now, the length $L$ of the cavity is $q$ times
the wavelength $\lambda$, with $q$ an integer. Then, setting the origin of the
$x$-axis in the center of the cavity, we have $x_j \in [-q\lambda/2,q\lambda/2]$, so that on using the periodicity of the cosine function, we can take the phase variables $\theta_j$ modulo $q$ such that $\theta_j \in [-\pi,\pi]$. Then, by measuring lengths in units of the reciprocal wavenumber $k^{-1} = \lambda/(2\pi)$ of the cavity standing wave, masses in units of the mass of the atoms $m$, and energies in units of $\hbar\Delta_c$, the Hamiltonian can be rewritten in dimensionless form as
\be
\mathcal{H} =  \sum_{j=1}^{N}\frac{(p_\theta)_j^{2}}{2} - \overline{n} N \Theta^2\,,
\label{eq:H-dl}
\ee
where, in terms of the $\theta$ variables, $\Theta$ is now expressed as
\be
\Theta = \frac{1}{N}\sum_{j = 1}^N \cos\theta_j\,.
\label{eq:phi}
\ee
The $(p_\theta)_j$'s in Eq.\ (\ref{eq:H-dl}) are 
the momenta canonically conjugated to the $\theta_j$ variables.

The similarity between the system with Hamiltonian (\ref{eq:H-dl}) and
an F-HMF model, already noted in \cite{Schutz:2014}, is now well
apparent, since the expression of the interaction field $\Theta$ given in Eq.\ (\ref{eq:phi}) coincides with the expression of the $x$-component of the magnetization of the HMF model. Indeed, the equations of motion derived from the Hamiltonian (\ref{eq:H-dl}) read
\be
\left\{
\begin{array}{ccl}
{\displaystyle \frac{{\rm d}\theta_j}{{\rm d}t}}& = & (p_\theta)_j\,,\\
& & \\ 
{\displaystyle \frac{{\rm d}(p_\theta)_j}{{\rm d}t}}& = & - \overline{n}\, \Theta \sin \theta_j\,, 
\end{array}
\right.
\label{eq:Morigi-EOM} 
\ee
for $j = 1,\ldots,N$. Equations (\ref{eq:Morigi-EOM}) coincide with
Eqs.\  (\ref{eq:EOM}) once $h = 0$, $\Theta = m_x$, $m_y = 0$ and $J =
\overline{n}$. Hence, the dynamics of a system of atoms interacting with
light in a cavity in the dissipationless limit is equivalent to that of
a model that differs from the ferromagnetic HMF model in zero field just
for the fact that particles in the former interact only with the
$x$-component of the magnetization. Equivalently, the dynamics is
equivalent to that of an HMF model with $m_y \equiv 0$. This means that the thermal equilibrium distribution for atoms in a cavity (still assuming that dissipative effects can be neglected) is still given by Eq.\ (\ref{eq:HMF-Gaussian-solution}), with $h = 0$ and $m_y = 0$.

\subsubsection{Quenching the coupling constant}
\label{sec:Morigi-qJ}

\begin{figure}
\begin{center}
\centering \includegraphics[width=10cm]{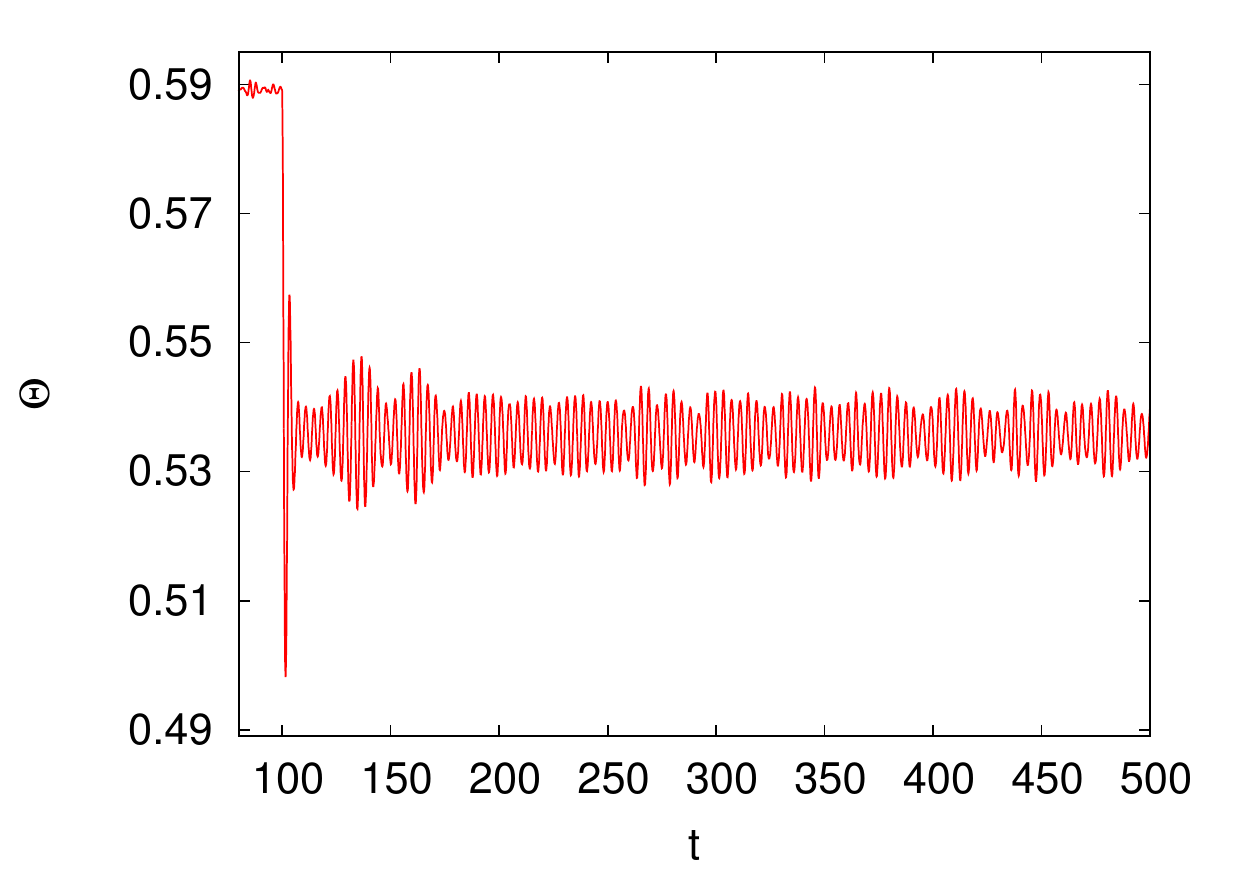}
\end{center}
\caption{Atoms in a cavity in the dissipationless limit. Time evolution
of the mean field $\Theta$: Starting with thermal equilibrium
(\ref{eq:HMF-Gaussian-solution}) at temperature $T=2$ and coupling
constant $\overline{n}=5$, the coupling is instantaneously
quenched at $t=100$ to $\overline{n}=4$. The number of particles is $N=10^6$.}
\label{fig:morigi-quench-J-mag}
\end{figure}

\begin{figure}
\begin{center}
\centering \includegraphics[width=10cm]{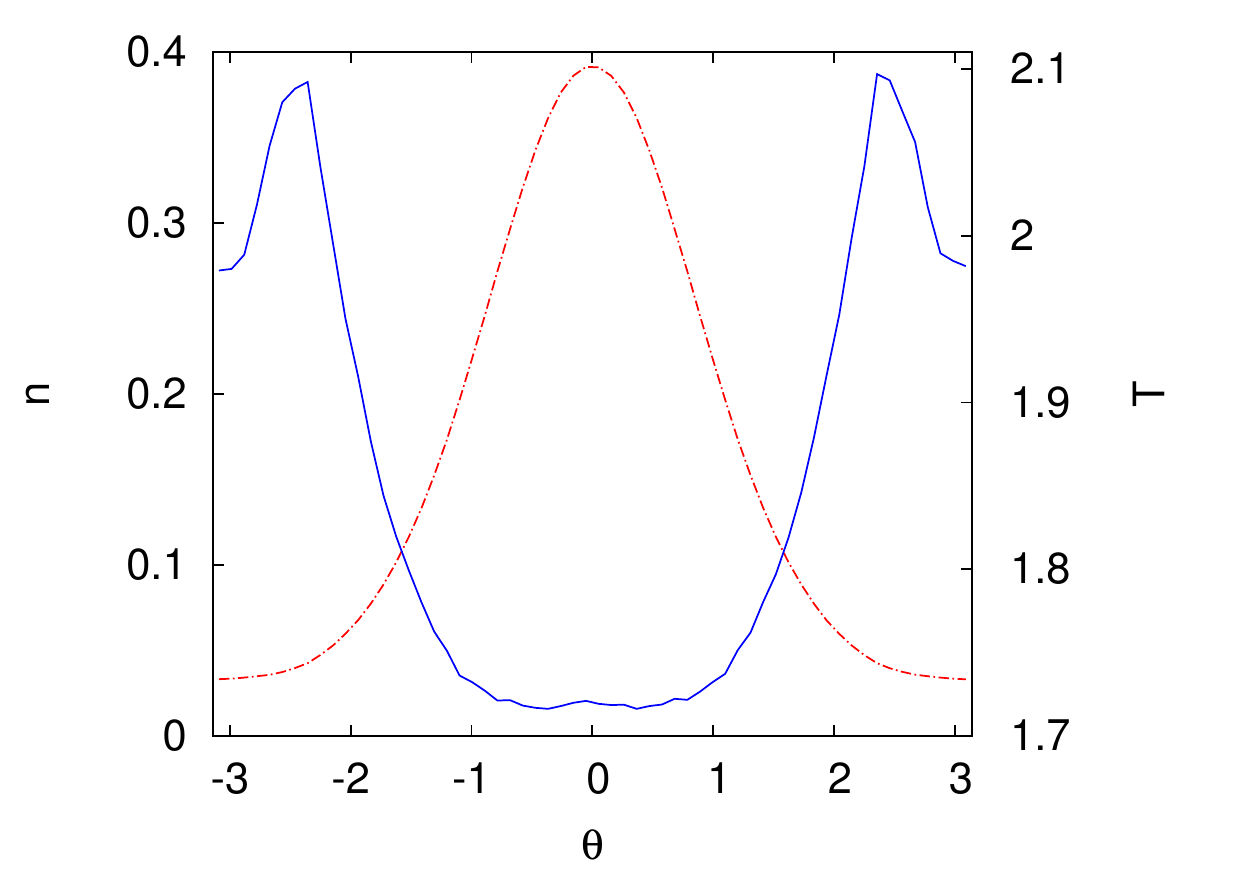}
\end{center}
\caption{Temperature inversion in a system of atoms in a cavity after a quench of the coupling. 
Local density $n$ (red dot-dashed line) and local temperature $T$ (blue
solid line) measured at $t = 10^3$ in the QSS obtained under the same
conditions as for Fig.\ \ref{fig:morigi-quench-J-mag}.}
\label{fig:morigi-quench-J-temp-inv}
\end{figure}

\begin{figure}
\centering \includegraphics[width=18cm]{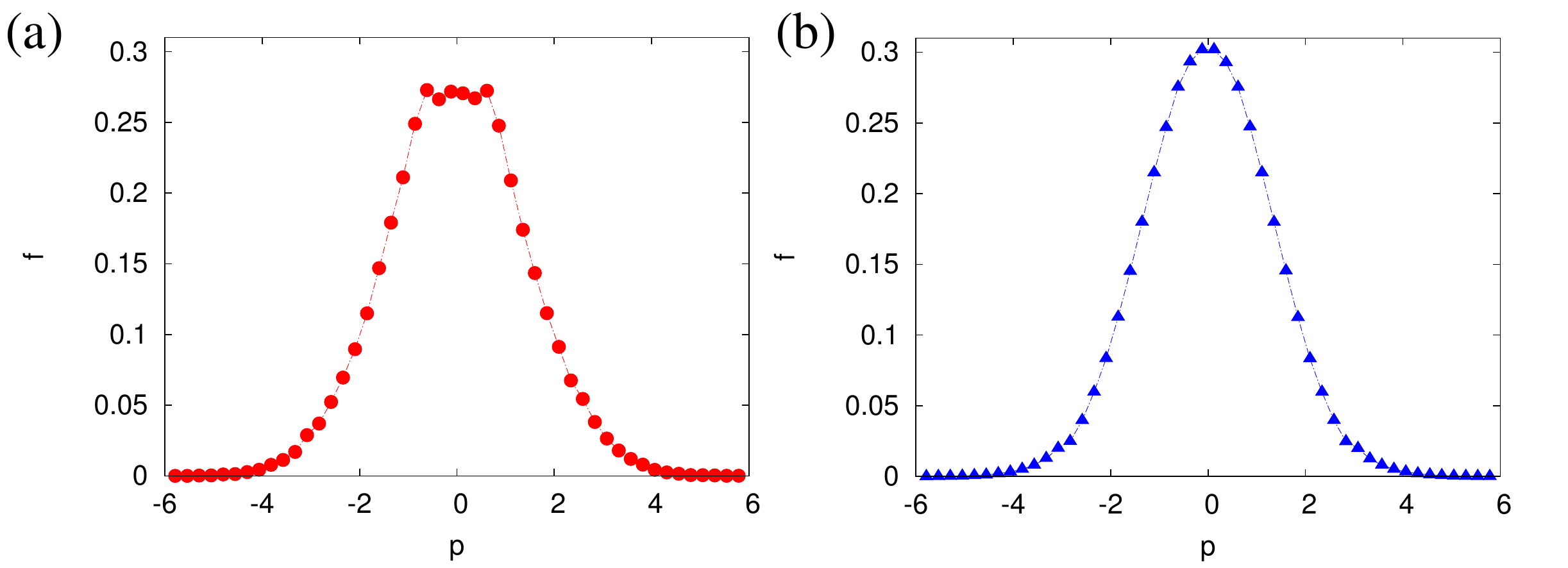}
\caption{ Momentum distribution of a system of atoms in a cavity measured at $t = 10^3$ in the QSS obtained under the same conditions as for Fig.\
\ref{fig:morigi-quench-J-mag}. (a) The distribution at $\theta=\pi$ at
which the density is minimum. (b) The distribution at $\theta=0$ at
which the density is maximum.}
\label{fig:morigi-quench-J-f(p)}
\end{figure}

In an HMF model in the thermodynamic limit $N\to\infty$, the condition
$m_y \equiv 0$ holds for all times if it holds at $t = 0$, that is, if
the initial spatial distribution is symmetric around $\theta = 0$. In a
finite system, even starting from a symmetric distribution, $m_y$ will
not stay exactly zero for all times, due to fluctuations induced by
finite-size effects, but will remain very small. Hence, the numerical
experiment reported in Sec.\ \ref{sec:F-qJ}, which was performed while
starting from an equilibrium state with $m_y(t = 0) = 0$, almost
directly applies also to the model of atoms in an optical cavity (with
$J = \overline{n}$), but for the fact that there was a coupling of the
particles to the small fluctuating $y$-component of the magnetization in
the former that would be absent in the atom case. It is reasonable to
expect that this coupling has only a small effect, since $N$ is very
large ($N = 10^6$). To check this, we repeated the numerical experiment
using the dynamics given by Eqs.\ (\ref{eq:Morigi-EOM}), that is, we
prepared the system in the same initial condition at $T = 2$ and
$\overline{n} = 5$, then evolved the dynamics---now using Eqs.\
(\ref{eq:Morigi-EOM})---until $t = 100$ when we quenched the coupling to
the new value $\overline{n} = 4$. The results are shown in Figs.\
\ref{fig:morigi-quench-J-mag}, \ref{fig:morigi-quench-J-temp-inv} and
\ref{fig:morigi-quench-J-f(p)}, that are essentially indistinguishable
from Figs.\ \ref{fig:hmf-quench-J-mag}, \ref{fig:hmf-quench-J-temp-inv}
and \ref{fig:hmf-quench-J-f(p)} of Sec.\ \ref{sec:F-qJ}. There is clear
temperature inversion, the momentum distribution exhibits the same
features as before, and the average temperature has the value $\langle T \rangle \approx 1.76$ in
 the QSS, lower
 than the initial value of $2$.  

\subsubsection{From numerical to laboratory experiments?}
\label{sec:Morigi-exp}

What do the results we have shown suggest as regards the possibility of
observing temperature inversion in a laboratory experiment with atoms
interacting with light in an optical cavity? To answer this question,
let us consider the system and the values of the parameters that were
already considered in Ref.\ \cite{Schutz:2014} so that the semiclassical
approximation for the atomic dynamics holds. We thus consider a system
of $^{85}{\rm Rb}$ atoms, whose mass is $m \approx 1.4 \times 10^{-25}$
kg; the atomic transition is the $D_2$ line, with a wavelength
$\lambda_0 = 780$ nm.  In terms of $\gamma=2\pi\times 3$ MHz, the
half-width of the $D_2$ line, the cavity linewidth is $\kappa=0.5\gamma$
and the detuning is $\Delta_c=-\kappa$. The energy unit is thus $\hbar
\Delta_c \approx 10^{-27}$ J. Preparing a system of atoms in a thermal
state with a temperature of order unity in these energy units means
reaching temperatures of the order of $10^{-4}$ K, which can be achieved
with laser cooling techniques. Hence, the initial state we have
considered in our simulations can be prepared in a laboratory, and, as
already noted, $\overline{n}$ can be tuned in the range we considered (and even in a
much wider range, if needed) so that also the quench of $\overline{n}$
is feasible. The crucial point is to understand whether the
dissipationless limit corresponding to assuming a Hamiltonian dynamics
reasonably describes the dynamics of the atoms over the timescale needed for the relaxation of the system to the QSS with temperature inversion. With our choice of energy, length and mass units, the time unit is fixed to
\be
\tau = \sqrt{\frac{m \lambda_0^2}{4\pi^2 \hbar \Delta_c}} \approx 10^{-6}~{\rm s},
\ee
this has to be compared with the timescale that rules the dissipative effects,
$\tau_c = \kappa^{-1} \approx 10^{-7}$ s. Luckily enough, the
simulations reported in Refs.\ \cite{Schutz:2014,Schutz:2015} show that
the dissipative effects set in on a timescale that is of the order of
$(10^4 \div 10^6)\tau_c$, while our simulations show that the QSS with
temperature inversion is reached after times of the order of $(10^2 \div
10^3)\tau_c$, so that there
should be ample room for measurements of temperature inversion before it is destroyed by dissipation.

\section{Iterative cooling via temperature inversion}
\label{sec:cooling}

We have already noted in all the cases considered above that the system
(be it an HMF or an atomic system) is colder in the QSS with temperature
inversion than it was in the initial thermal equilibrium state.
Moreover, the fact that the QSS has temperature inversion means that
hotter particles reside in the external parts of the system, where the
density is smaller. This suggests that temperature inversion could be
exploited to make the system even colder by means of an iterative
protocol that goes as follows. We prepare the system in an inhomogeneous
thermal state at temperature $T$ symmetric about $\theta = 0$, perform a
quench, and let the system relax to the QSS with temperature inversion
as before. Now, we filter out hotter particles, that is, all the particles that are located at $|\theta| > \theta_c$, with a suitable choice of $\theta_c$: the resulting system will be close to isothermal at a temperature $T'< T$. We let the system relax for some time and then perform another quench. The system goes into another QSS with temperature inversion and average temperature $T'' < T'$. The protocol can be iterated as long as it continues to appreciably cool the system, or as long as there is room for another quench of the parameters.

We now show the results obtained by applying this iterative cooling protocol to the cases studied before. 

\subsection{Cooling the antiferromagnetic HMF model}

\begin{figure}[!h]
\includegraphics[width=18cm]{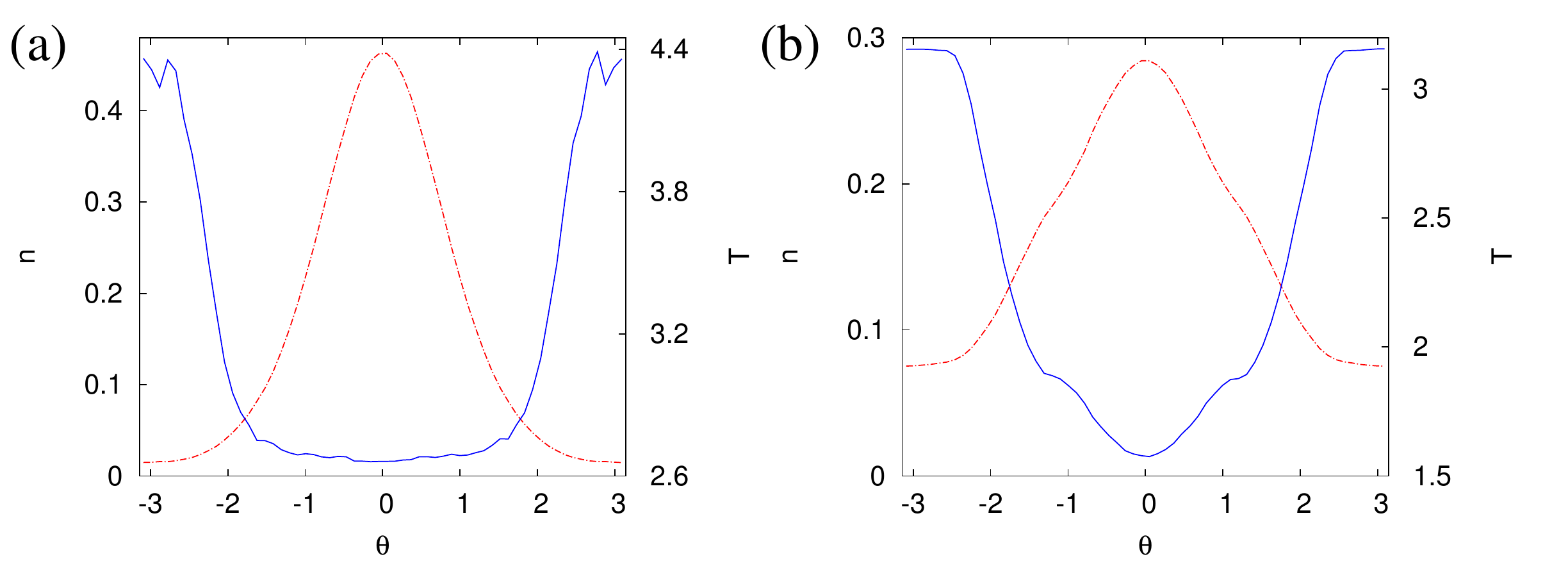}
\caption{Iterative cooling in the AF-HMF model in presence
of an external field. (a) Local density $n$ (red dashed-dotted line)
and local temperature $T$ (blue solid line) in the QSS after the first filtering stage of the hot particles. (b) Local density $n$ (red dashed-dotted line)
and local temperature $T$ (blue solid line) in the QSS after the second
filtering stage. $N$ and $J$ values are the same as in Fig.\ \ref{fig:ahmf-quench-h-temp-inv}.}
\label{fig:ahmf-cooling-dens-temp}
\end{figure}

\begin{figure}[!h]
\centering
\includegraphics[width=10cm]{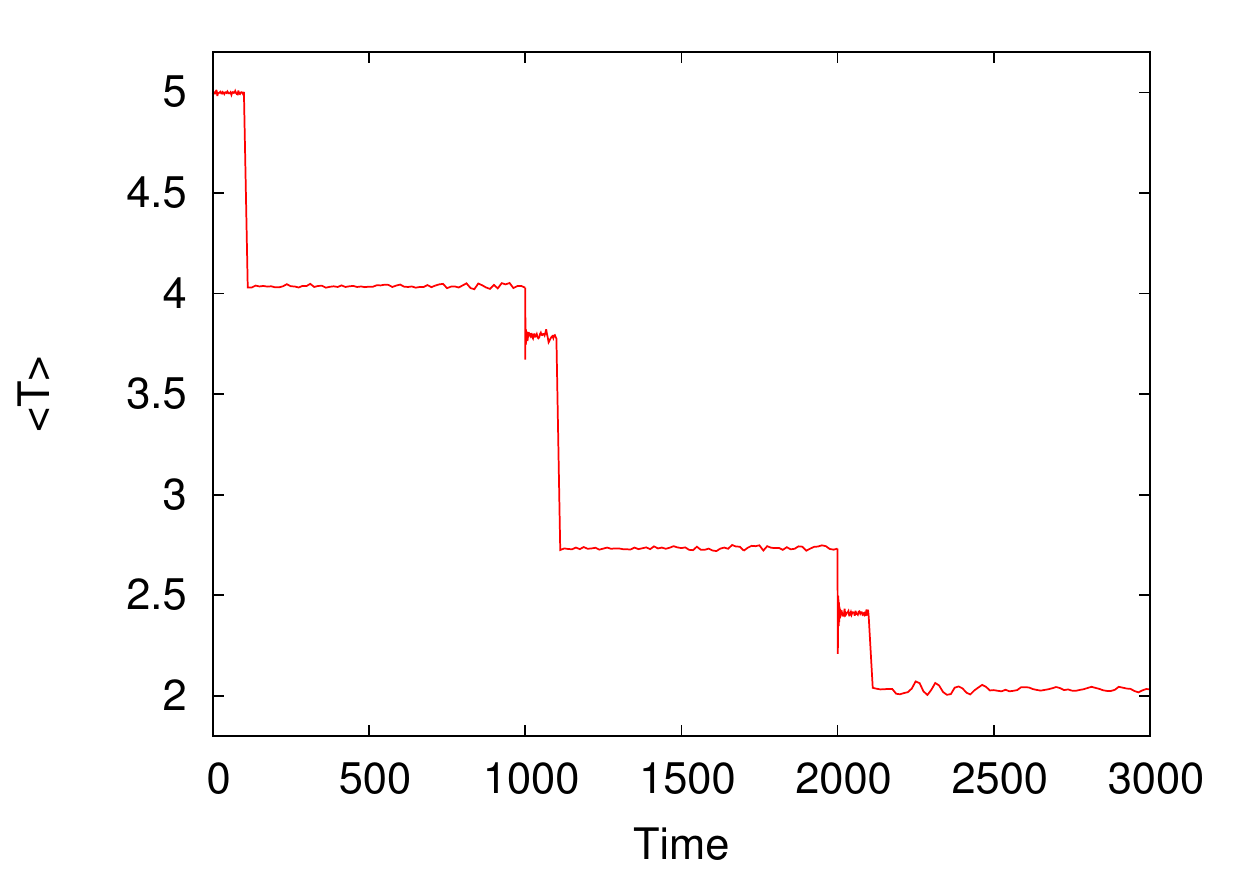}
\caption{Iterative cooling in the AF-HMF model in presence
of an external field. Average temperature $\langle T \rangle$ as a
function of time for the full cooling protocol. $N$ and $J$ values are
the same as in Fig.\ \ref{fig:ahmf-quench-h-temp-inv}.}
\label{fig:ahmf-cooling-average}
\end{figure}

We start with the AF-HMF model. After performing the quench described in
Sec.\ \ref{sec:AF-qh}, whereby starting in equilibrium at $J = -1$, $T =
5$ and $h = 15$, the field was quenched to $h = 10$, the system is in a
QSS with temperature and density profiles shown in Fig.\
\ref{fig:ahmf-quench-h-temp-inv}. Now, we filter out the hotter
(high-energy) particles, which we choose to be those with $\theta$'s
lying outside the interval $[-2,2]$. Subsequently, we instantaneously quench the field from $h = 10$ to
$h=5$; the system eventually settles into a QSS, and the corresponding local density and local temperature profiles are shown in Fig.\ \ref{fig:ahmf-cooling-dens-temp} (a). By further filtering out the high-energy particles and quenching instantaneously the field to
$h=1.0$, the system goes to another QSS whose density and temperature profiles are reported in Fig.\ \ref{fig:ahmf-cooling-dens-temp} (b). The average temperature as a
function of time for the full sequence of events starting with equilibrium at $T = 5$ and $h = 15$ is shown in Fig. \ref{fig:ahmf-cooling-average}. The system cools down from an initial average temperature $\langle T \rangle=5$ to a final value $\langle T \rangle\approx 2$. Each small downward step in $\langle T \rangle(t)$ corresponds to
filtering out the high-energy particles, while each large downward
step corresponds to settling of the system into a QSS with a lower temperature as a result
of quenching of the field to a lower value. The percentage decrease in the number of particles during filtering out the high-energy particles equals about $2.11\%$ during the first stage of filtration, and about $4.95\%$ during the second stage. 
Not surprisingly, the cooling protocol becomes less efficient at subsequent stages.

\subsection{Cooling the ferromagnetic HMF model and atoms in a cavity}

\begin{figure}[!h]
\includegraphics[width=18cm]{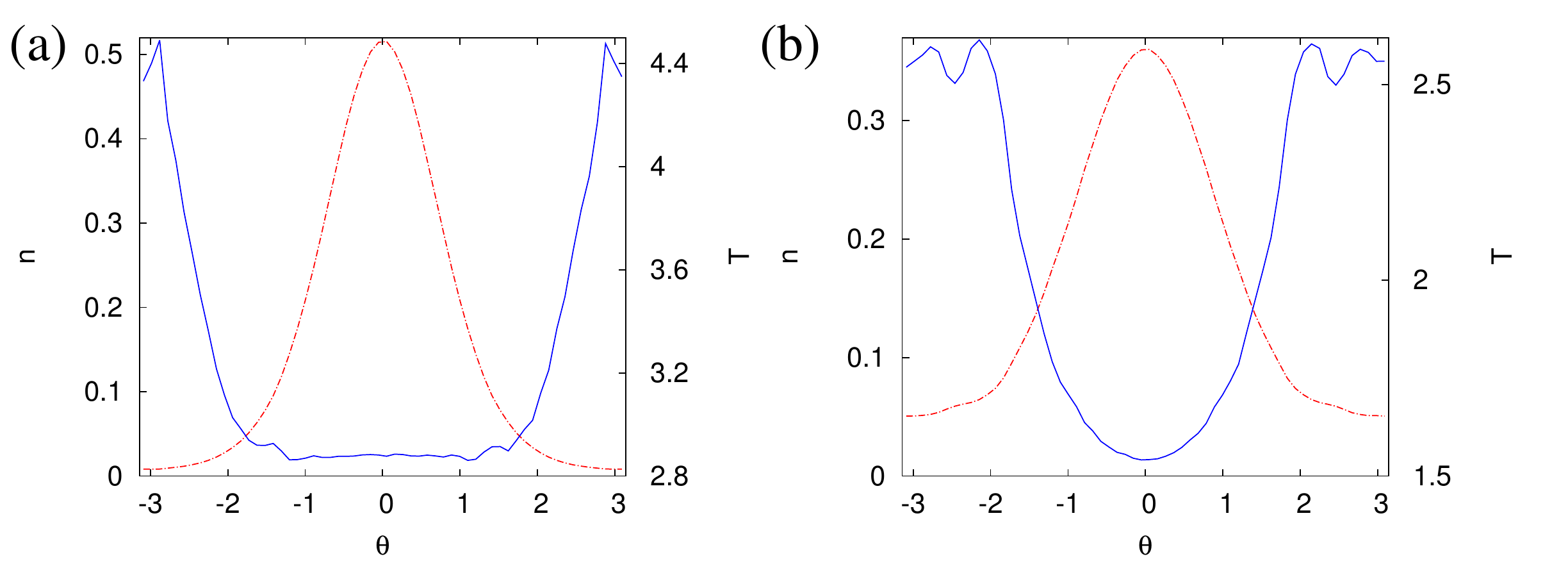}
\caption{Iterative cooling in the F-HMF model in presence
of an external field. (a) Local density $n$ (red dashed-dotted line)
and local temperature $T$ (blue solid line) in the QSS after the first filtering stage of the hot particles. (b) Local density $n$ (red dashed-dotted line)
and local temperature $T$ (blue solid line) in the QSS after the second
filtering stage. $N$ and $J$ values are the same as in Fig.\ \ref{fig:hmf-quench-h-temp-inv}.}
\label{fig:hmf-cooling-dens-temp}
\end{figure}

\begin{figure}[!h]
\centering
\includegraphics[width=10cm]{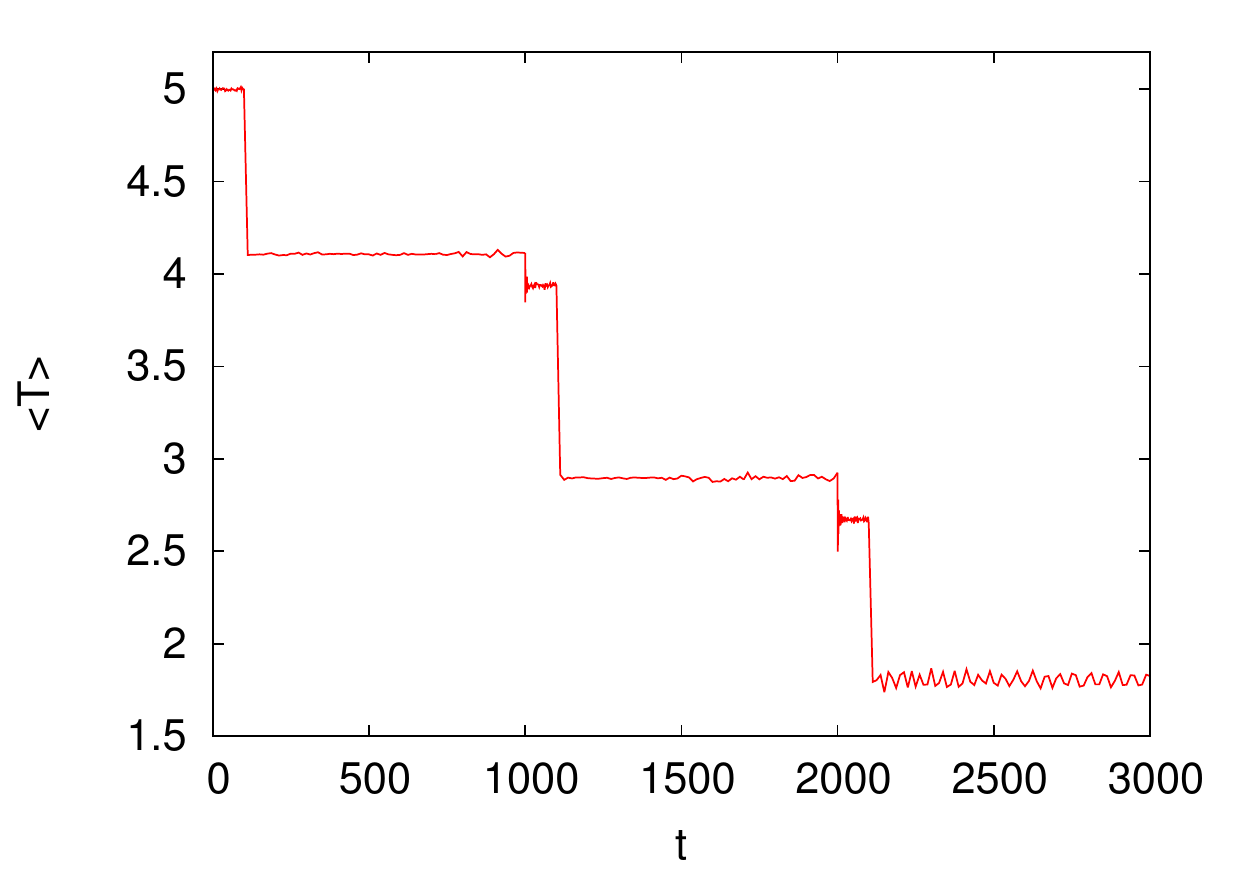}
\caption{Iterative cooling in the F-HMF model in presence
of an external field. Average temperature $\langle T \rangle$ as a
function of time for the full cooling protocol. $N$ and $J$ values are
the same as in Fig.\ \ref{fig:hmf-quench-h-temp-inv}.}
\label{fig:hmf-cooling-average}
\end{figure}

\begin{figure}[!h]
\centering
\includegraphics[width=10cm]{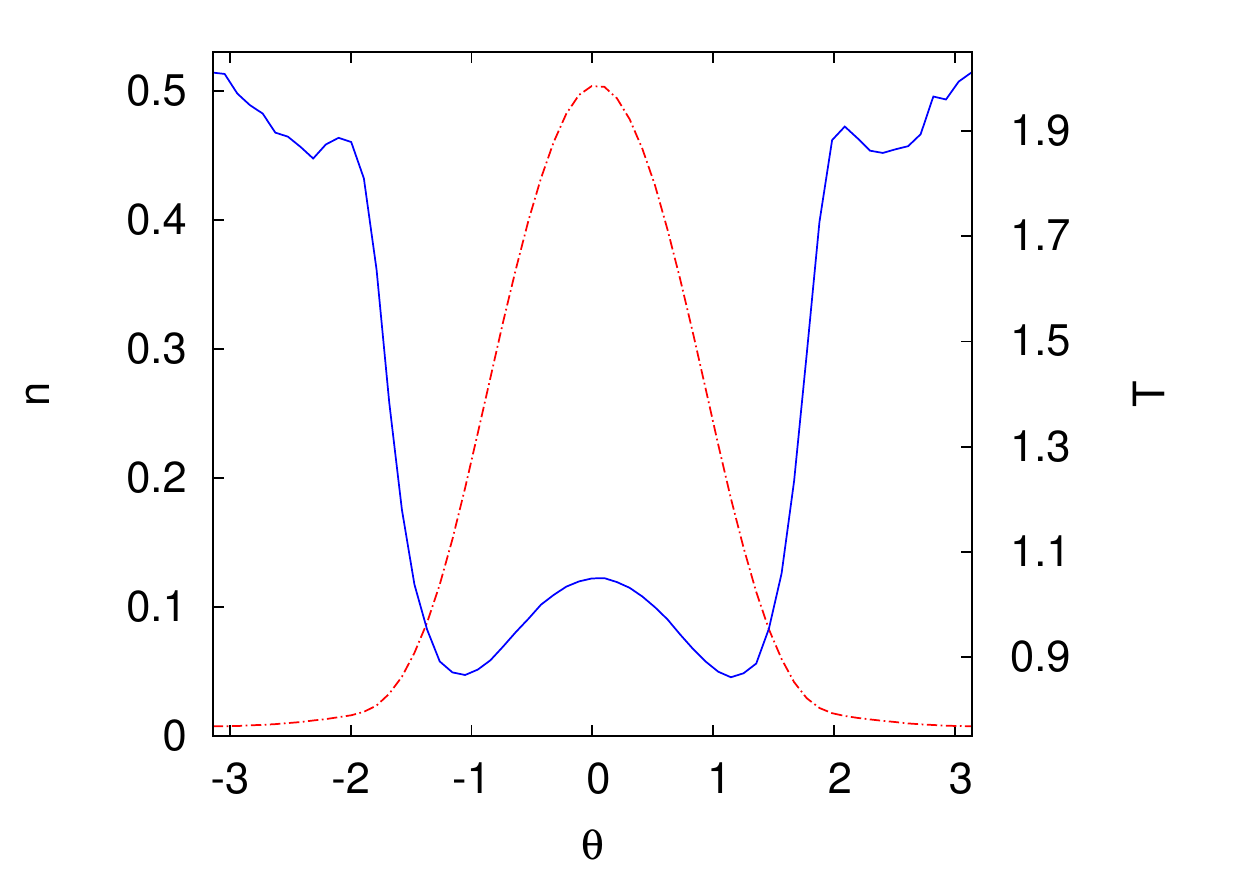}
\caption{Iterative cooling in the F-HMF model without external field: local density $n$ (red dashed-dotted line)
and local temperature $T$ (blue solid line) in the QSS after the
filtering stage of the hot particles. $N$ value is the same as in Fig.\ \ref{fig:hmf-quench-J-temp-inv}.}
\label{fig:hmf-cooling-J-dens-temp}
\end{figure}

\begin{figure}[!h]
\centering
\includegraphics[width=10cm]{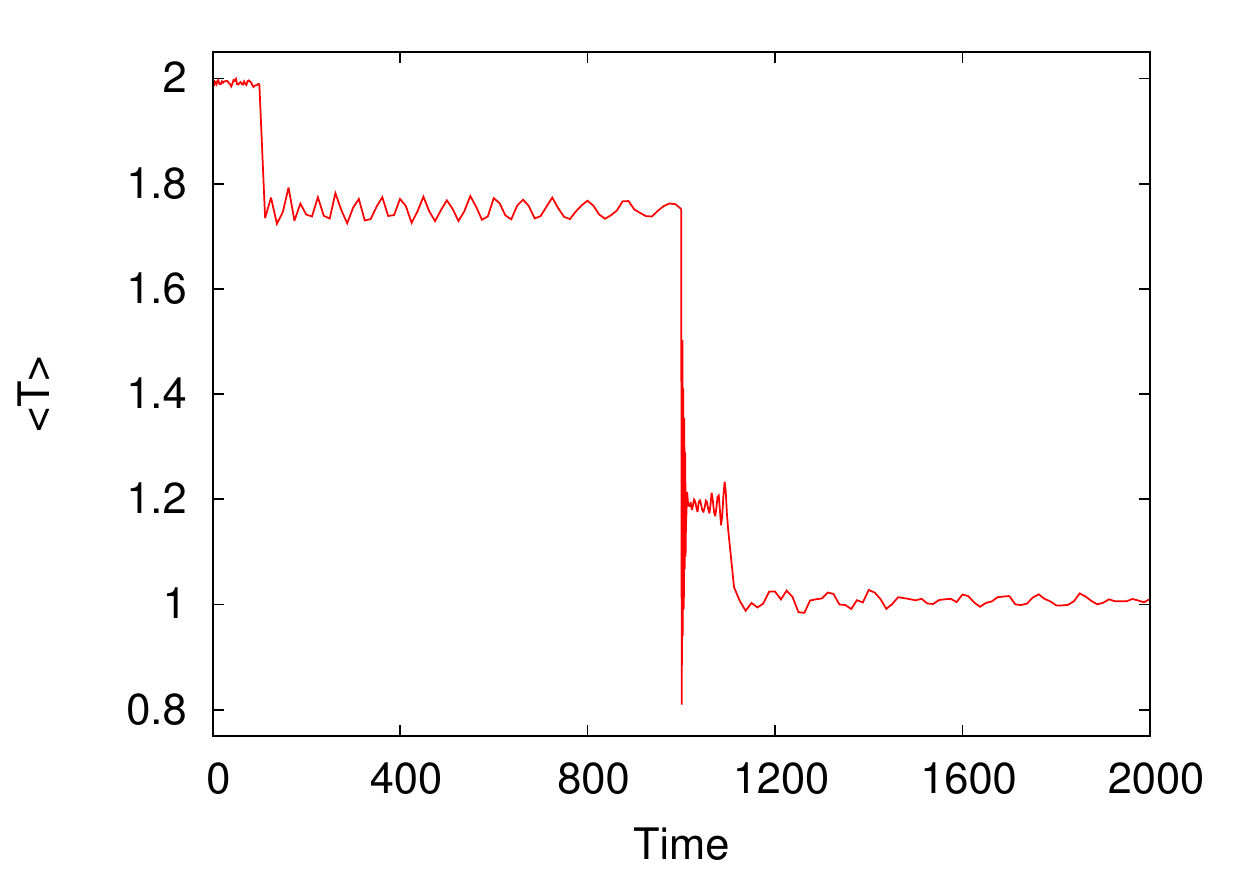}
\caption{Iterative cooling in the F-HMF model without an external field. Average temperature $\langle T \rangle$ as a
function of time for the full cooling protocol. $N$ value is the same as in Fig.\ \ref{fig:hmf-quench-J-temp-inv}.}
\label{fig:hmf-cooling-J-average}
\end{figure}

We now apply the same cooling protocol described above to the F-HMF
model in an external field, while starting from equilibrium  at $J = 1$,
$T = 5$ and $h = 15$. The results are
shown in Figs.\ \ref{fig:hmf-cooling-dens-temp} and \ref{fig:hmf-cooling-average}. Again, we obtain a cooling from an initial average temperature $\langle T \rangle=5$ to a final value $\langle T \rangle\approx 2$. The percentage decrease in the number of particles during filtering out the high-energy particles equals 
 about $1.24\%$ during the first stage of filtration, and equals about
 $3.16\%$ during the second stage.
We now turn to the case of a quench of the coupling constant, still for
the F-HMF model but now without external field. The cooling protocol
goes as before, but we can fruitfully perform only one stage of
filtering instead of two. We start with the equilibrium state at $T = 2$
and $J = 5$ that is quenched to $J = 4$, yielding the QSS whose density
and temperature profiles are shown in Fig.\
\ref{fig:hmf-quench-J-temp-inv}. We then filter out the hotter
particles, in this case defined as those with $\theta$'s lying outside
the interval $[-1,1]$. Subsequently, we instantaneously quench the
coupling to $J=3$. The density and temperature profile in the resulting
QSS are shown in Fig.\ \ref{fig:hmf-cooling-J-dens-temp}, and the time
evolution of the average temperature is reported in Fig.\
\ref{fig:hmf-cooling-J-average}. Under the full quenching protocol, the system
cools down from an initial average temperature $\langle T \rangle=2$
to a final value $\langle T \rangle\approx 1.0$. The percentage decrease in the
number of particles during filtering out the high-energy particles
equals about $34.3\%$.


\begin{figure}[!h]
\centering
\includegraphics[width=10cm]{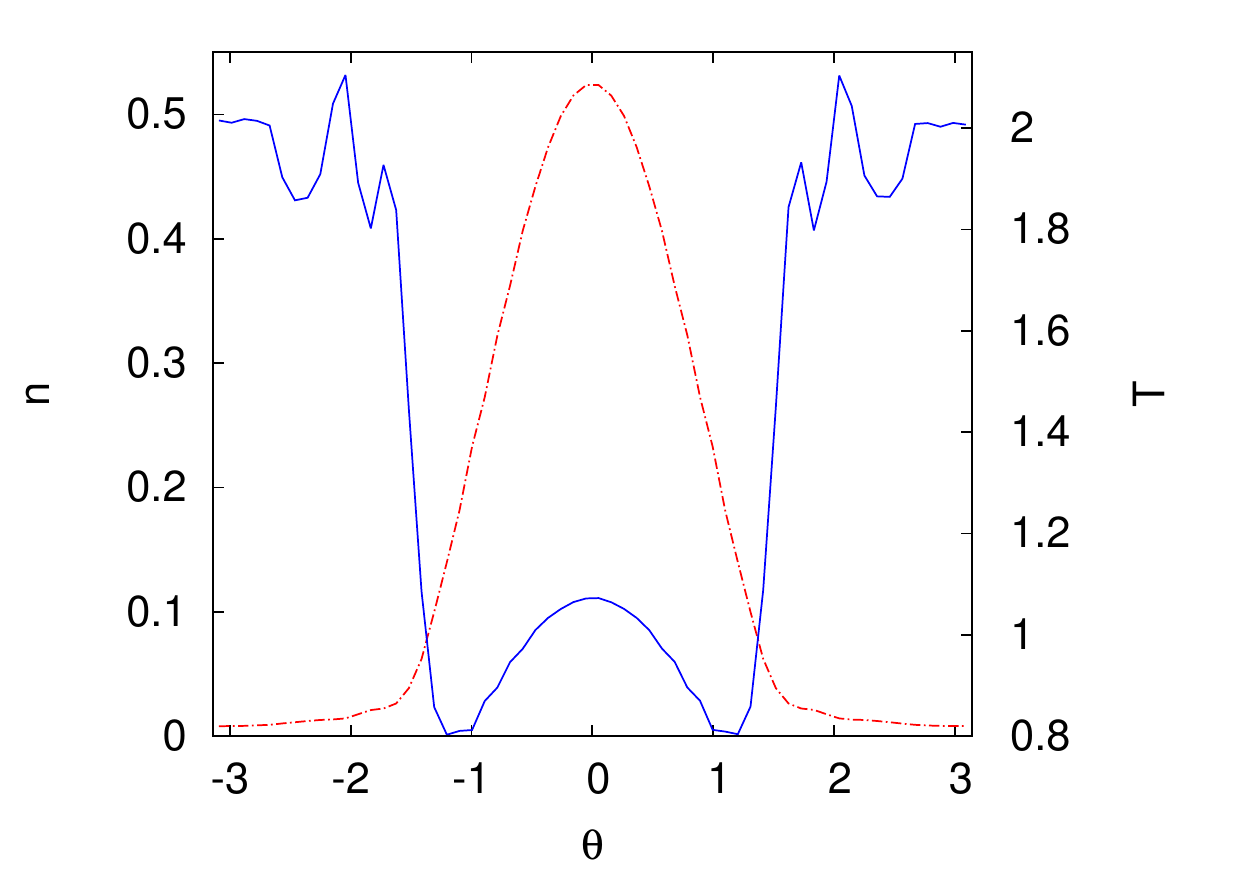}
\caption{Iterative cooling of atoms in a cavity: local density $n$ (red
dashed-dotted line) and local temperature $T$ (blue solid line) in the
QSS after the filtering stage of the hot particles. The value of $N$ is
the same as in Fig.\ \ref{fig:morigi-quench-J-temp-inv}.}
\label{fig:morigi-cooling-J-dens-temp}
\end{figure}

\begin{figure}[!h]
\centering
\includegraphics[width=10cm]{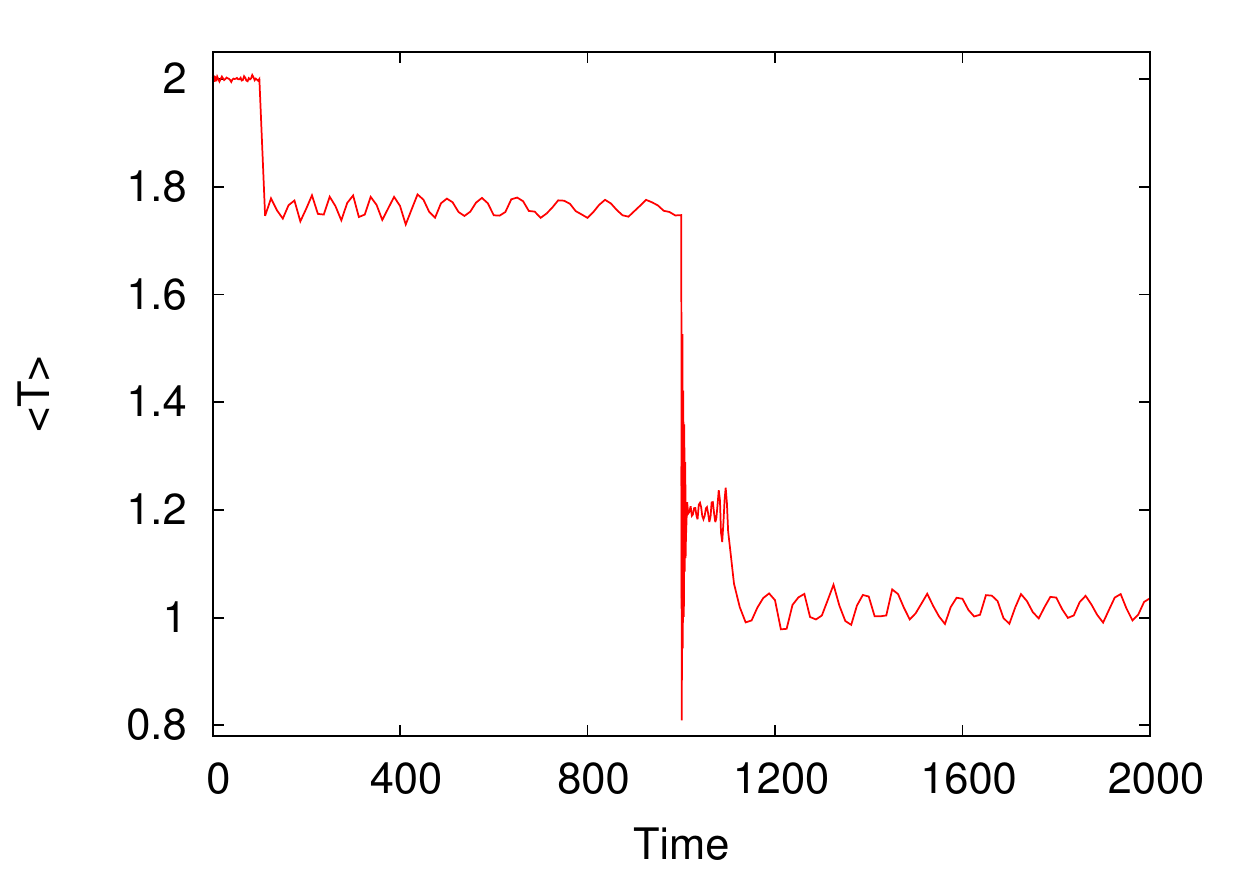}
\caption{Iterative cooling of atoms in a cavity: average temperature
$\langle T \rangle$ as a function of time for the full cooling protocol.
$N$ value is the same as in Fig.\ \ref{fig:morigi-quench-J-temp-inv}.}
\label{fig:morigi-cooling-J-average}
\end{figure}

Finally, we apply exactly the same cooling protocol to the dynamics of atoms in a cavity, given by Eqs.\ \ref{eq:Morigi-EOM}. The results are shown in Figs.\ \ref{fig:morigi-cooling-J-dens-temp} and \ref{fig:morigi-cooling-J-average}, and as expected are very close to that obtained for the F-HMF model: the system under the quenching protocol
cools down, from an initial average temperature $\langle T \rangle=2$
to a final value $\langle T \rangle\approx 1$. The percentage decrease in the number of particles during filtering out the high-energy particles equals about $34.5\%$.

Although the cooling protocol does not yield dramatic results,
especially in the last two considered cases, still it seems able to (at
least) decrease the temperature of a system by a factor of two.
Moreover, it opens up the possibility of a different way of cooling a system with respect to the nowadays commonly used ones.

\section{Concluding remarks}
\label{sec:conclusions}
In this paper, we have demonstrated that quasi-stationary states that an
isolated long-range interacting system relaxes to when starting from a
spatially inhomogeneous thermal equilibrium and subsequently subject to
a quench of one of the parameters of the Hamiltonian generically exhibit
the phenomenon of temperature inversion. Using molecular dynamics
simulations of a prototypical model, the HMF model, we have shown that
different quench protocols lead to temperature inversion, in presence
and in absence of an external field, and with both attractive and
repulsive interactions: no fine-tuning is necessary in any of the cases.
This lends further support to the claim made in
\cite{Teles:2015,Casetti:2014} that, rather than a peculiarity of some
astrophysical systems, temperature inversion should be a ``universal''
feature of nonequilibrium quasi-stationary states resulting from
perturbations of spatially inhomogeneous thermal equilibrium. Moreover,
our analysis of the momentum distribution functions supports the
interpretation put forward in \cite{Teles:2015} that temperature
inversion arises due to an interplay between wave-particle interaction
(the role of the wave being played by the oscillation of the mean field
induced by the perturbation or the quench) and velocity filtration. The
latter is a mechanism proposed by Scudder
\cite{Scudder:1992-1,Scudder:1992-2,Scudder:1994} to explain the
temperature profile of the solar corona, and basically amounts to the
statement that a suprathermal velocity distribution becomes broader when climbing a potential well.

The fact that quench protocols are able to produce temperature
inversions is of particular relevance in view of observing this
phenomenon in controlled laboratory experiments. Exploiting the close
connection between the HMF model and a model describing the dynamics of
a system of atoms interacting with light in an optical cavity, as noted
in Ref.\ \cite{Schutz:2014}, we have argued that a quench of the
external laser pump should be able to produce a quasi-stationary state
with temperature inversion in a system of, say, Rubidium atoms in a
cavity that is initially in thermal equilibrium at temperatures
reachable by laser cooling. Such a quasi-stationary state should survive
for quite a long time before being destroyed by dissipative effects. The
latter claim is made on the assumption that the timescales of
dissipation are the same as those observed in
\cite{Schutz:2014,Schutz:2015}. Similar to our study, these works also
considered a quench of the coupling constant, but from different initial
states and to different final states, and this may affect the
dissipation timescales. However, results in \cite{Schutz:2015} seem to
indicate that dissipation is inhibited in spatially inhomogeneous
states, and this would be an advantage for a quench experiment like
those suggested here. An interesting follow-up of the present work, in
view of a detailed study of the feasibility of an experiment, would
surely be to use the techniques of \cite{Schutz:2014,Schutz:2015} to
study the evolution after quenches like the ones described here under a
dynamics that fully takes into account the dissipative effects. Another
interesting system to study in view of a possible experiment to detect
temperature inversion would be a system of trapped ions, which may be
related to the antiferromagnetic HMF model. Work is in progress along
this direction \cite{Landa:2016}.

Finally, we have shown that temperature inversion may be exploited to
cool the system. The cooling protocol is of iterative type, and takes
advantage of the fact that in states with temperature inversion hotter
particles are spatially separated from colder ones, so that they can be
filtered out. Let us sign off by saying that temperature inversion is
one of many fascinating features of the nonequilibrium states of
long-range interacting systems that are yet to be unveiled.

\section{Acknowledgments}
We thank P.\ Di Cintio, H.\ Landa and A.\ Trombettoni for fruitful discussions and suggestions. SG acknowledges the support and hospitality of INFN (Italy) and the University of Florence.
\vspace{1cm}

\end{document}